\documentclass{aa}  
\usepackage[flushleft]{threeparttable}
\usepackage{float}
\usepackage{graphics}
\usepackage{graphicx}
\usepackage{amssymb, amsmath}
\usepackage{tikz}
\usetikzlibrary{graphs, positioning, quotes, shapes.geometric}
\usepackage{amstext}
\usepackage[T1]{fontenc}
\usepackage{mathrsfs}
\usepackage[normalem]{ulem}
\usepackage{natbib,twoopt}
\usepackage{hyperref} 
\usepackage{soul,xcolor}
\usepackage{booktabs}
\usepackage{rotating}
\usepackage{txfonts}
\makeatletter
\def\@to{to}
\def\as     {\ifmmode {\rlap.}$\,$''$\,$\! \else ${\rlap.}$\,$''$\,$\!$\fi}
     \def\decsec  {\ifmmode {\rlap.}$\,$^{\rm s}$\,$\! \else ${\rlap.}$\,$^{\rm s}$\,$\!$\fi}\def\decss  {\ifmmode {\rlap.}$\,$^{\rm s}$\,$\! \else ${\rlap.}$\,$^{\rm s}$\,$\!$\fi}
\makeatother

\newcommand{\rah}{$^{\mbox{\scriptsize h}}$}
\newcommand{\ram}{$^{\mbox{\scriptsize m}}$}
\newcommand{\ras}{$^{\mbox{\scriptsize s}}$}
\newcommand{\decd}{$^{\circ}$}
\newcommand{\decm}{$'$}
\newcommand{\decs}{$\farcs$}

\newcommand{\choh}{CH$_{3}$OH}

\newcommand{\ch}{C$_{3}$H$_{2}$}

\begin{document}

   \title{Multi-line observations of CH$_{3}$OH, c-C$_{3}$H$_{2}$ and HNCO towards L1544}

   \subtitle{Dissecting the core structure with chemical differentiation}

   \author{Y. Lin         
          \inst{1}
          \and S. Spezzano
          \inst{1}
          \and O. Sipil{\"a}
          \inst{1}
          \and A. Vasyunin
          \inst{2}
          \and P. Caselli\inst{1}}

   \institute{Max-Planck-Institut f{\"u}r Extraterrestrische Physik, Giessenbachstr. 1, D-85748 Garching bei M{\"u}nchen\\
              \email{ylin@mpe.mpg.de}
         \and
            Ural Federal University, 620002, 19 Mira street, Yekaterinburg, Russia \\
             }

   \date{Received ; accepted }

 
  \abstract
   {Pre-stellar cores are the basic unit for the formation of stars and stellar systems. The anatomy of the physical and chemical structures of pre-stellar cores is critical for understanding the star formation process.}
   {L1544 is a prototypical pre-stellar core, which shows significant chemical differentiation surrounding the dust peak. We aim to constrain the physical conditions at the different molecular emission peaks. This study allows us to compare the abundance profiles predicted from chemical models together with the classical density structure of Bonnor-Ebert (BE) sphere.}
   {We conducted multi-transition pointed observations of CH$_{3}$OH, c-C$_{3}$H$_{2}$ and HNCO with the IRAM 30m telescope, towards the dust peak and the respective molecular peaks of L1544. Using this data set, with non-LTE radiative transfer calculations and a 1-dimensional model, we revisit the physical structure of L1544, and benchmark with the abundance profiles from current chemical models.}
   {We find that the HNCO, c-C$_{3}$H$_{2}$ and CH$_{3}$OH lines in L1544 are tracing progressively higher density gas, from $\sim$10$^{4}$ to several times 10$^{5}$ cm$^{-3}$. Particularly, we find that to produce the observed intensities and ratios of the CH$_{3}$OH lines, a local gas density enhancement upon the BE sphere is required. This suggests that the physical structure of an early-stage core may not necessarily follow a smooth decrease of gas density profile locally, but can be intercepted by clumpy substructures surrounding the gravitational center. }
   {Multiple transitions of molecular lines from different molecular species can provide a tomographic view of the density structure of pre-stellar cores. The local gas density enhancement deviating from the BE sphere may reflect the impact of accretion flows that appear asymmetric and are enhanced at the meeting point of large-scale cloud structures.}

   \maketitle
%

\section{Introduction}

Dense cores, with hydrogen gas densities of $\gtrsim$10$^{4}$ cm$^{-3}$ and physical scale of 0.1 pc, are the birth places of stars and stellar systems (\citealt{BT07},\citealt{diFrancesco07}). Pre-stellar cores represent an early phase of core evolution after the starless core, which appear gravitationally bound but still absent of protostars. Understanding physical structures of pre-stellar cores can provide important constraints on the initial condition for the formation of low-mass stars. During the core evolution, in accordance with variations of physical structures, chemical compositions can vary significantly inside the core (\citealt{Suzuki92}, \citealt{VD98}, \citealt{Ceccarelli07}, \citealt{BT07}). Furthermore, also the environment has an effect on the chemical structure of the embedded core (\citealt{Spezzano16, Spezzano20}). 
A combination of multiple transitions of different molecular species provides an indispensable diagnosis kit for analysing both the past and current physical properties of dense cores. 

Gravitational collapse of dense cores leads to the formation of low-mass stars, but the detailed process of the collapse remains elusive. One of the widely used models is the Bonnor-Ebert (BE) sphere (\citealt{KC10}, \citealt{BK10}). Essentially, the model suggests that cores are hydrostatic objects during gravitational contraction, which are balanced by the thermal pressure complemented with the turbulence kinetic energy (\citealt{Nakano98}). Subsequent core evolution is caused by turbulence damping during which the core remains in quasi-static equilibrium. Additional support from magnetic field allows for an increment of the central gas density of the core (\citealt{KetoField05}). The density profile of a BE sphere has a plateau in the center followed by a $r^{-2}$ decrease in outer regions, and the scale of the central plateau decreases as the collapse proceeds (\citealt{KC10}). In contrast to this quasi-static model, the turbulent origin suggests that core formation is driven by compressive gas motions, a more dynamical process (\citealt{MK04}). Pre-stellar cores are pristine sites to distinguish between the competing core formation models, and they also provide unique test beds for understanding the complex chemistry of molecular gas. 

Recent ALMA observations of the innermost regions of pre-stellar cores have resolved or tentatively detected small-scale ($\lesssim$1000 au-0.01 pc) substructures (e.g., \citealt{Ohashi18}, \citealt{Caselli19}, \citealt{Tokuda20}, \citealt{Sahu21}). While the origin of the substructures is still under debate, it suggests a BE sphere description of pre-stellar cores may be over-simplified except for the featureless cores. The substructures can also exhibit large chemical differences (e.g., \citealt{Tatematsu20}), suggesting that they are at different evolutionary stages. It is also uncertain whether these observed substructures will coalesce in a later stage or collapse individually (\citealt{Sahu21}). In general, turbulent fragmentation predicts a high level of multiplicity inside the core (\citealt{Goodwin04}, \citealt{Offner10}), but how fragmentation starts and evolves remains unclear. Investigations on the physical structure of pre-stellar cores, in which gravitational contraction has just started to critically shape the core gas, can provide important constraints on this aspect.



L1544 is a well-studied late-stage pre-stellar core. Its physical structure seems to be well described by the BE sphere (\citealt{KC10}, \citealt{Keto14, Keto15}, \citealt{Caselli22}). Inside the core, gravitational collapse has started, showing infall motions, but the central singularity has not yet formed (\citealt{Myers96}, \citealt{Tafalla98}, \citealt{Caselli02b}, \citealt{KC10}, \citealt{Caselli19, Caselli22}). The chemical composition of L1544 exhibits significant spatial differentiation: the inner core is depicted by high deuterium fraction (\citealt{Caselli02b, Caselli02a}, \citealt{Crapsi07}, \citealt{CT19b}, \citealt{R19}); surrounding the dust peak ($\gtrsim$0.02 pc) at different positions, the core displays characteristic emission peaks of molecules (\citealt{Spezzano16, Spezzano17}, \citealt{JS16}). The three molecular peaks are the CH$_{3}$OH peak, the c-C$_{3}$H$_{2}$ peak and the HNCO peak, which also appear to be emission peaks for several other chemically related species of the three molecules, respectively (\citealt{Spezzano17}). The physical and chemical mechanisms behind the marked chemical differentiation are not fully known. The non-uniform illumination by the large-scale radiation field seems to play an important role resulting in the distinct emission peaks of CH$_{3}$OH and c-C$_{3}$H$_{2}$, by influencing the amount of gas-phase atomic carbon; this is verified towards a sample of pre-stellar cores that show different spatial distribution of CH$_{3}$OH and c-C$_{3}$H$_{2}$ emission (\citealt{Spezzano20}). 

Because chemical processes are tightly connected with physical properties, probing the excitation conditions of these molecules can provide important clues on their chemical differentiation (\citealt{VD98}, \citealt{CC12}, \citealt{Jorgensen20}). To this end, we have carried out multi-line observations of CH$_{3}$OH, c-C$_{3}$H$_{2}$ and HNCO towards L1544 (Table \ref{tab:lines_more}), targeting at the dust peak and the molecular peaks (Table \ref{tab:coord}). 
By incorporating the abundance profiles predicted by chemical models (\citealt{Sipila16}, \citealt{vasyunin17}) in full radiative transfer calculations with spherical symmetry, we gauge how well the current chemical models can reproduce the observed line emission.  

The paper is laid out as follows: in Sec. \ref{sec:obs} information of observations and data reduction procedure are presented. The obtained molecular lines are described in Sec. \ref{sec:sps_basic}. In Sec. \ref{sec:radex} and \ref{sec:loc} calculations of radiative transfer models are elaborated, providing constraints on the physical properties. Lastly, results from different radiative transfer models and from different molecular species are discussed in Sec. \ref{sec:dis}, and our conclusion and outlook are presented in Sec. \ref{sec:conclusion}.

\begin{figure*}[htb]
  \hspace{.25cm}\includegraphics[scale=0.3]{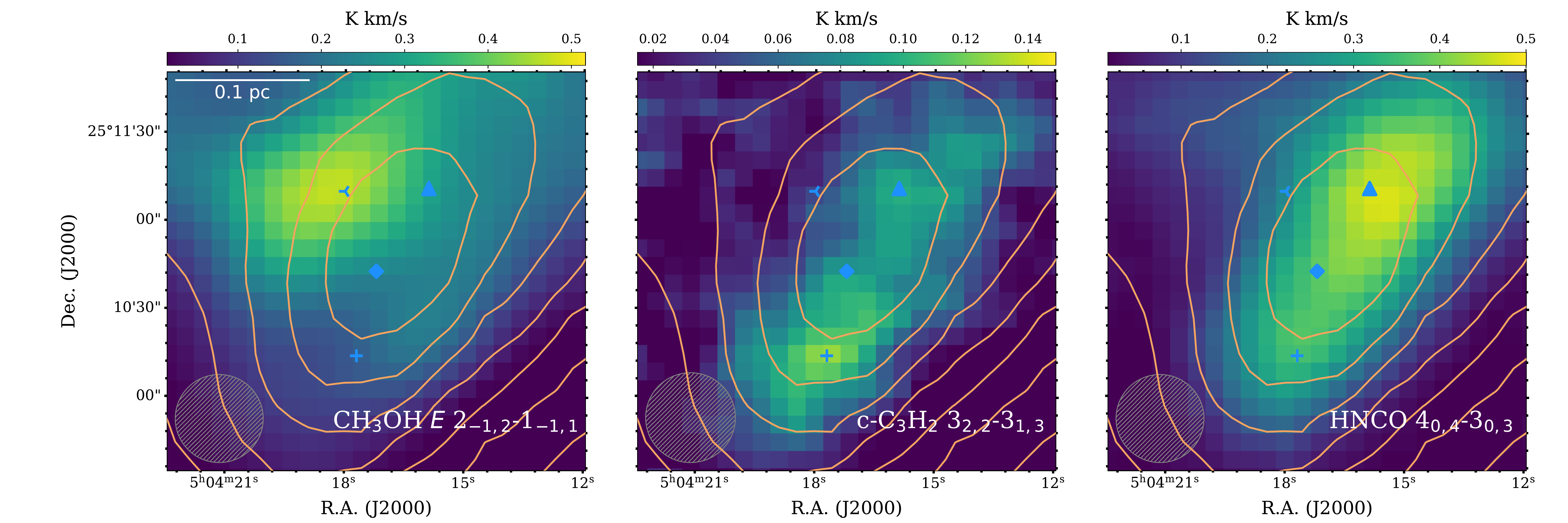}
  \caption{The integrated intensity maps (in colorscale) of CH$_{3}$OH-$E$ 2$_{-1, 2}$-1$_{-1, 1}$, c-C$_{3}$H$_{2}$ 3$_{2, 2}$-3$_{1, 3}$ and HNCO 4$_{0, 4}$-3$_{0, 3}$ (from left to right) observed by the IRAM 30m telescope, showing the different locations of the molecular peaks (Table \ref{tab:coord}, \citealt{Spezzano17}). The contour levels (in orange) show 3.2$\times$10$^{21}$, 4.3$\times$10$^{21}$, 5.9$\times$10$^{21}$, 8.1$\times$10$^{21}$, 1.1$\times$10$^{22}$, 1.5$\times$10$^{22}$, and 2.1$\times$10$^{22}$ cm$^{-2}$ of the molecular hydrogen column density ($N_{\mathrm{H_{2}}}$) map (\citealt{Spezzano16}). The markers in blue represent the dust peak (diamond), molecular emission peaks of CH$_{3}$OH (three-branched triangle),  HNCO (upper triangle) and c-C$_{3}$H$_{2}$ (plus).}
  \label{fig:guidemaps}
\end{figure*}

\begin{table*}
\centering
\begin{threeparttable}
\caption{Observed lines.}
\label{tab:lines_more}
\begin{tabular}{lccccc}
\toprule
 Molecule&Transitions  & Frequency & $E_{\mathrm{up}}$ &Critical density $^{a}$& 30m Beam\\
  &&(GHz)&(K)& ($\times$10$^{4}$\,cm$^{-3}$) &(HPFW in $''$)\\
  \midrule
  CH$_{3}$OH-$E$&2$_{-1,2}$-1$_{-1,1}$ &  96.7394 & 12.5 &  1.3 &  26.0 \\
   &2$_{0,2}$-1$_{0,1}$ &  96.7445 & 20.1 & 15.4 &  26.0 \\
   &2$_{1,2}$-1$_{1,1}$ &  96.7555 & 28.0 & 31.5 &  26.0 \\
   &3$_{0,3}$-2$_{0,2}$ & 145.0938 & 27.1 & 26.2 &  17.3 \\
 &3$_{-1,3}$-2$_{-1,2}$ & 145.0974 & 19.5 &  4.4 &  17.3 \\
   &3$_{1,3}$-2$_{1,2}$ & 145.1319 & 35.0 & 46.1 &  17.3 \\
  &4$_{0,4}$-4$_{-1,4}$ & 157.2461 & 36.3 & 44.2 &  16.0 \\
  &1$_{0,1}$-1$_{-1,1}$ & 157.2708 & 15.4 &  9.3 &  16.0 \\
  &2$_{0,2}$-2$_{-1,2}$ & 157.2760 & 20.1 & 15.3 &  16.0 \\
   &3$_{0,3}$-3$_{-1,3}$ & 157.2723 & 27.1 & 26.2 &  16.0 \\

  c-C$_{3}$H$_{2}$   &3$_{1,2}$-2$_{2,1}$ & 145.0896 & 16.1 & 10.4 &  17.3 \\
   &3$_{3,0}$-2$_{2,1}$ & 216.2788 & 19.5 & 33.7 &  11.6 \\
   &6$_{1,6}$-5$_{0,5}$ & 217.8221 & 38.6 & 70.8 &  11.5 \\
   &4$_{3,2}$-3$_{2,1}$ & 227.1691 & 29.1 & 52.0 &  11.1 \\
   
  HNCO$^{b}$&5$_{0,5}$-4$_{0,4}$ & 109.9058 & 15.8 &  5.1 &  22.9 \\
   &6$_{0,6}$-5$_{0,5}$ & 131.8857 & 22.1 &  8.8 &  19.1 \\
   &7$_{0,7}$-6$_{0,6}$ & 153.8651 & 29.5 & 13.9 &  16.3 \\
   &8$_{0,8}$-7$_{0,7}$ & 175.8437 & 38.0 & 21.5 &  14.3 \\ 
  

  \bottomrule
\end{tabular}
    \begin{tablenotes}
      \small
      \item The laboratory spectroscopy references for CH$_{3}$OH is \citet{Xu08}, for c-C$_{3}$H$_{2}$ is \citet{Vrtilek87}
and for HNCO \citet{Hocking75}.
      \item a: Calculated following definition in \citet{Shirley15} in the optically thin limit at 10 K, considering a multi-level energy system whenever necessary. 
     \item b: Para-H$_{2}$ is assumed to be the collision partner.
      \end{tablenotes}
  \end{threeparttable}
\end{table*}

%

\section{Observations}\label{sec:obs}
Observations of the molecular lines listed in Table \ref{tab:lines_more} were taken with the IRAM 30m telescope. All the lines, except \choh\,2$_{K}$-1$_{K}$, are 
observed with pointed observations towards the dust peak and the molecular peaks of L1544. These observations were conducted on January 28-29th, March 1st and April 17th, 2021 (Project: 101-20, PI: Silvia Spezzano) using EMIR with FTS backend and a tracked frequency switch mode. The typical precipitable water vapor is 2-3 mm. The spectral resolution is 0.10 km s$^{-1}$ at 145 GHz, and the corresponding beam size $\sim$17$''$ (half-power full width, here after HPFW, listed in Table \ref{tab:lines_more}). The achieved rms level (1$\sigma$) is $\sim$15-25 mK. Typical calibration uncertainties are 20$\%$.
We used the CLASS software package in Gildas for the data reduction. A first-order baseline subtraction was applied. The antenna temperatures ($T_{\mathrm{A}}^{\star}$) were converted to main-beam brightness temperature ($T_{\mathrm{mb}}$) with efficiencies interpolated according to the online $\eta_{\mathrm{mb}}$ table \footnote{https://publicwiki.iram.es/Iram30mEfficiencies} for different line frequencies. 

We also adopted the archival IRAM 30m telescope mapping observations of \choh\,(2-1) lines of L1544, some of which were previously published by \citet{Bizzocchi14} and \citet{Spezzano16}.  

\begin{table}[htb]
\centering
\begin{threeparttable}
\caption{Coordinates of the dust peak and three molecular peaks.}
\label{tab:coord}
\begin{tabular}{lcc}
\toprule
Location & R.A.&Dec.\\
&(J2000)&(J2000)\\
  \midrule

Dust peak  &05\rah04\ram17\ras.21& 25\decd10\decm42\decs8\\
HNCO peak& 05\rah04\ram15\ras.90 &25\decd11\decm11\decs1\\
c-C$_{3}$H$_{2}$ peak&  05\rah04\ram17\ras.70& 25\decd10\decm14\decs0\\
CH$_{3}$OH peak& 05\rah04\ram18\ras.00& 25\decd11\decm10\decs0\\
 \bottomrule
\end{tabular}
  \end{threeparttable}
\end{table}

\begin{figure*}
    \centering
    \includegraphics[scale=0.35]{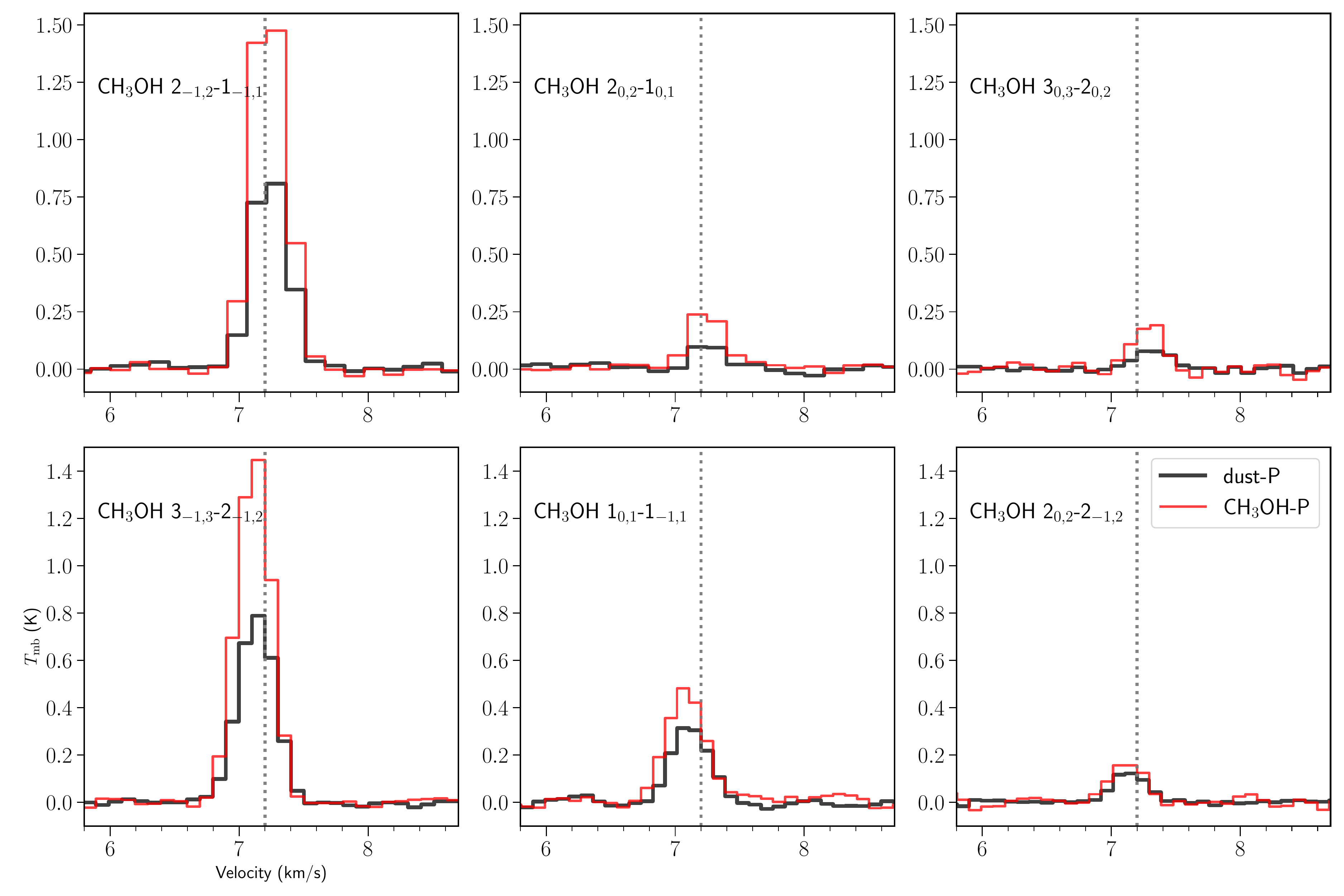}\\
    \includegraphics[scale=0.35]{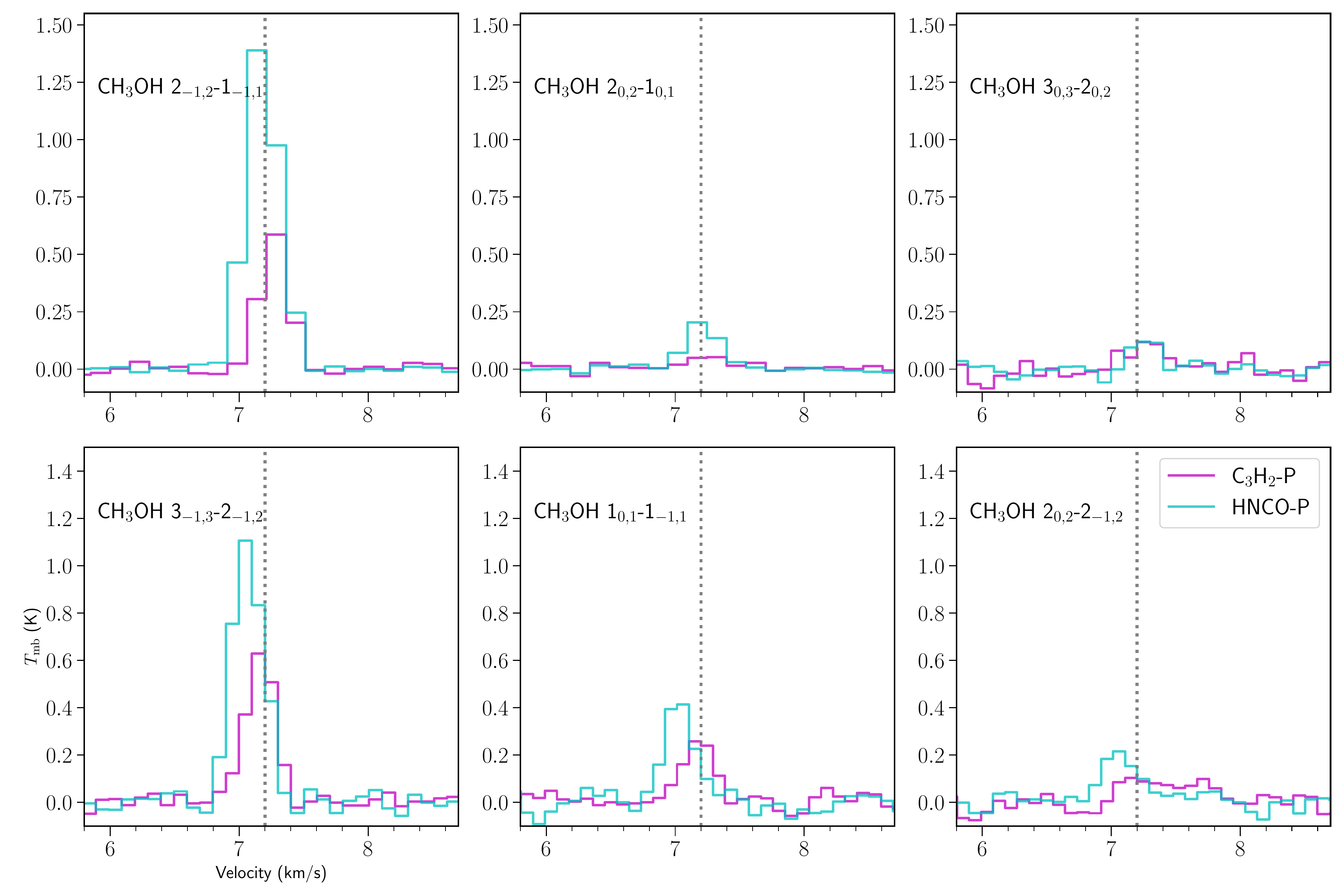}

    \caption{The observed CH$_{3}$OH spectra at the dust peak and at the molecular peaks. The spectra are arranged in order of increasing frequencies. Vertical dotted lines indicate $V_{\mathrm{LSR}}$ = 7.19 km s$^{-1}$.}
    \label{fig:sps_allwithdust}
\end{figure*}

\begin{figure*}
    \centering
        \includegraphics[scale=0.4]{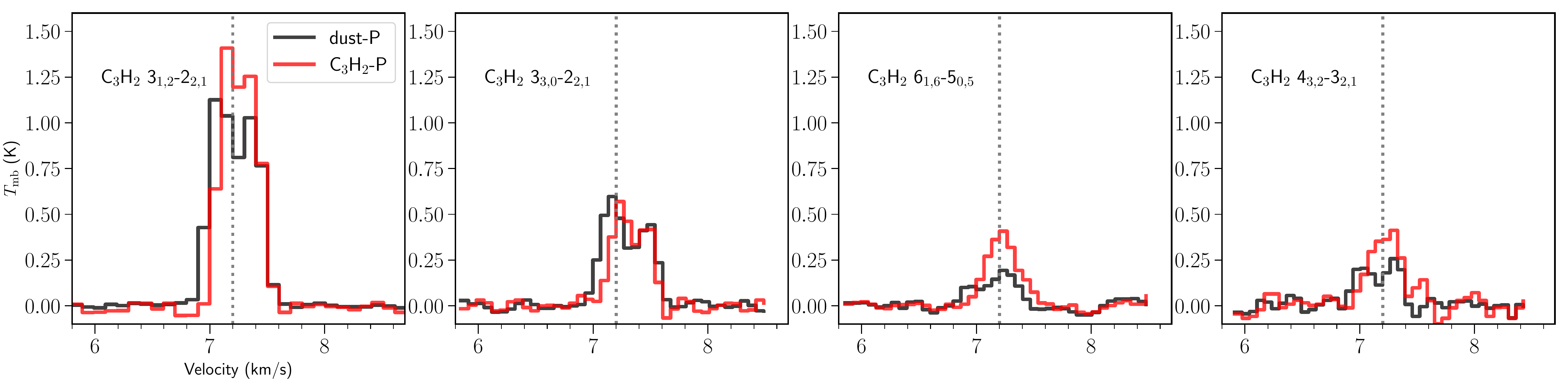}\\
        \includegraphics[scale=0.4]{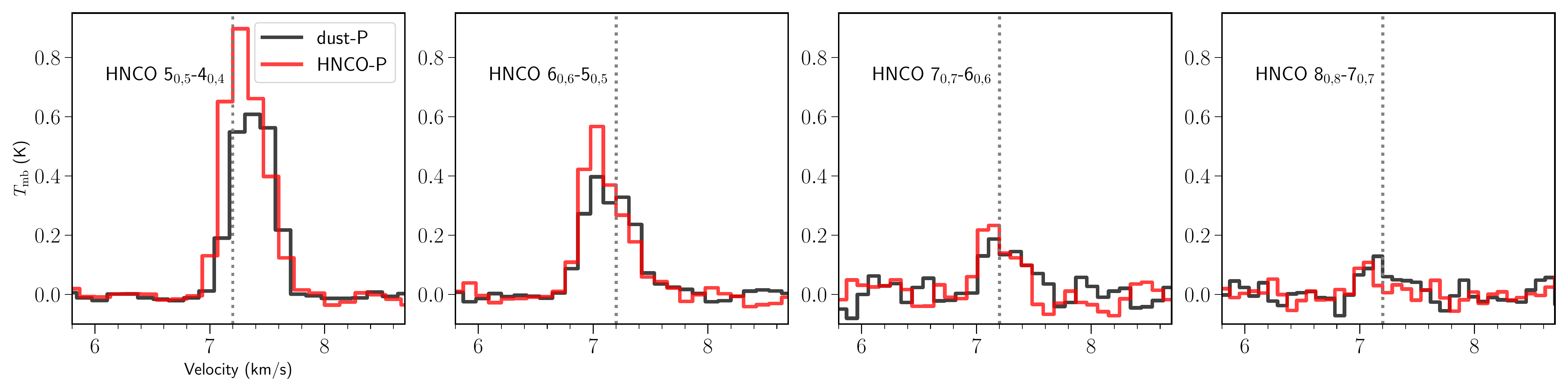}\\
\caption{The observed c-C$_{3}$H$_{2}$ and HNCO spectra at the dust peak and at the molecular peaks. The spectra are arranged in order of increasing frequencies. Vertical dotted lines indicate $V_{\mathrm{LSR}}$ = 7.19 km s$^{-1}$.}
    \label{fig:sps_allwithdust1}
\end{figure*}

\section{Results}
\subsection{Spectra at the dust peak and molecular peaks}\label{sec:sps_basic}

The obtained molecular lines at the dust peak and the molecular peaks are shown in Fig. \ref{fig:sps_allwithdust} and Fig. \ref{fig:sps_allwithdust1}. All the $E_{\mathrm{up}}$$\lesssim$20 K transitions of CH$_{3}$OH were detected above 5$\sigma$ towards the dust peak and all three molecular peaks. These lines have critical densities $\lesssim$1.5$\times$10$^{5}$ cm$^{-3}$. Towards the CH$_{3}$OH peak, additionally, the 3$_{0,3}$-2$_{0,2}$ line ($E_{\mathrm{up}}$$\sim$27 K, $n_{\mathrm{crit}}$$\sim$2.6$\times$10$^{5}$ cm$^{-3}$) was detected. 

The c-C$_{3}$H$_{2}$ and HNCO lines were only observed towards the dust peak and the respective molecular peaks. at the dust peak and c-C$_{3}$H$_{2}$ peak, all four lines of c-C$_{3}$H$_{2}$ in Table 1 were detected. The 4$_{3,2}$-3$_{2,1}$ and 3$_{3,0}$-2$_{2,1}$ lines at the dust peak show an emission dip at the $V_{\mathrm{LSR}}$ of L1544 (7.19 km s$^{-1}$). At the c-C$_{3}$H$_{2}$ peak, there is also a dip in the 3$_{3,0}$-2$_{2,1}$ line profile, but not in the 4$_{3,2}$-3$_{2,1}$ line. As for the four HNCO lines, at the dust peak and HNCO peak, all the lines were detected except the line of the highest energy level (8$_{0,8}$-7$_{0,7}$), and have a signal-to-noise ratio of $>$5 (Fig. \ref{fig:sps_allwithdust1}).  

We fitted the molecular line profiles with Gaussian models, the parameters of which are listed in Tables \ref{tab:gauss_lines} to \ref{tab:gauss_lines1}. The line widths ($\sigma_{\mathrm{V}}$) obtained are mostly $\sim$0.15$\pm$0.03 km s$^{-1}$, with no significant variations between lines of different molecular species or different positions.

\subsection{Radiative transfer modeling}\label{sec:sps_modeling}

\begin{table*}[htb]
\centering
\begin{threeparttable}
\caption{Gaussian profile parameters for lines at the c-C$_{3}$H$_{2}$ peak and the CH$_{3}$OH peak.}\label{tab:gauss_lines}
\begin{tabular}{lllllll}
\toprule
Position& \multicolumn{3}{l}{C$_{3}$H$_{2}$-P} & \multicolumn{3}{l}{CH$_{3}$OH-P} \\
Gauss parameter & $T_{\mathrm{mb}}$ &     $V_{\mathrm{LSR}}$ &  $\sigma_{V}$ & $T_{\mathrm{mb}}$ &     $V_{\mathrm{LSR}}$ &  $\sigma_{V}$  \\
            &  K   &         km s$^{-1}$          &    km s$^{-1}$             &       K      &       km s$^{-1}$              &    km s$^{-1}$            \\
\midrule
c-C$_{3}$H$_{2}$ 3$_{1, 2}$-2$_{2, 1}$ &        1.43(0.04) &  7.22(0.01) &  0.16(0.01) &        0.73(0.09) &  7.15(0.02) &  0.15(0.02) \\
c-C$_{3}$H$_{2}$ 3$_{3, 0}$-2$_{2, 1}$ &        0.53(0.03) &  7.22(0.01) &  0.16(0.01) &               N   &         N   &         N   \\
c-C$_{3}$H$_{2}$ 4$_{3, 2}$-3$_{2, 1}$ &        0.38(0.04) &  7.25(0.01) &  0.13(0.01) &               N   &         N   &         N   \\
c-C$_{3}$H$_{2}$ 6$_{0, 6}$-5$_{1, 5}$ &        0.40(0.02) &  7.25(0.01) &  0.14(0.01) &               N   &         N   &         N    \\
CH$_{3}$OH 1$_{0, 1}$-1$_{-1, 1}$     &        0.28(0.03) &  7.19(0.01) &  0.13(0.01) &        0.49(0.01) &  7.09(0.00) &  0.15(0.00)  \\
CH$_{3}$OH 2$_{0, 2}$-2$_{-1, 2}$     &                 $\lesssim$0.09 &           - &           - &        0.16(0.01) &  7.10(0.01) &  0.13(0.01) \\
CH$_{3}$OH 3$_{0, 3}$-2$_{0, 2}$    &                 $\lesssim$0.09 &           - &           - &        0.20(0.02) &  7.14(0.01) &  0.11(0.01) \\
CH$_{3}$OH 3$_{0, 3}$-3$_{-1, 3}$     &                 $\lesssim$0.09&           - &         -  &$\lesssim$0.04 &                 - &           -       \\
CH$_{3}$OH 3$_{1, 3}$-2$_{1, 2}$     &        0.65(0.02) &  7.20(0.00) &  0.12(0.00) &        1.47(0.01) &  7.15(0.00) &  0.13(0.00)  \\
CH$_{3}$OH 4$_{0, 4}$-4$_{-1, 4}$     &                 $\lesssim$0.10 &           - &           - &                 $\lesssim$0.05&           - &           -  \\
HNCO 5$_{0, 5}$-4$_{0, 4}$           &               N   &         N   &         N   &               N   &         N   &         N   \\
HNCO 6$_{0, 6}$-5$_{0, 5}$           &               N   &         N   &         N   &               N   &         N   &         N   \\
HNCO 7$_{0, 7}$-6$_{0, 6}$           &               N   &         N   &         N   &               N   &         N   &         N   \\
HNCO 8$_{0, 8}$-7$_{0, 7}$           &               N   &         N   &         N   &               N   &         N   &         N   \\
\bottomrule
\end{tabular}
\begin{tablenotes}
\item ``N'' means not observed; for lines not detected, we quote the 3$\sigma$ level as upper limits with other parameters denoted as ``-''.
\item The estimated standard error for each variable of Gaussian model is indicated in parentheses.
\end{tablenotes}
\end{threeparttable}
\end{table*}

\begin{table*}[htb]
\centering
\begin{threeparttable}
\caption{Gaussian profile parameters for lines at the dust peak and the HNCO peak.}\label{tab:gauss_lines1}

\begin{tabular}{lllllll}
\toprule
Position & \multicolumn{3}{l}{dust-P} & \multicolumn{3}{l}{HNCO-P} \\
Gauss parameter &  $T_{\mathrm{mb}}$ &                 $V_{\mathrm{LSR}}$ &              $\sigma_{V}$ & $T_{\mathrm{mb}}$ &     $V_{\mathrm{LSR}}$ &  $\sigma_{V}$ \\
              &   K   &         km s$^{-1}$          &    km s$^{-1}$             &       K      &       km s$^{-1}$              &    km s$^{-1}$       \\
\midrule
c-C$_{3}$H$_{2}$ 3$_{1, 2}$-2$_{2, 1}$ &   1.10(0.06) &  7.18(0.01) &  0.21(0.01) &        1.08(0.07) &  7.14(0.02) &  0.19(0.02) \\
c-C$_{3}$H$_{2}$ 3$_{3, 0}$-2$_{2, 1}$ & 0.60(0.03) &   7.20(0.02)  &   0.25(0.02) &               N   &         N   &         N   \\
c-C$_{3}$H$_{2}$ 4$_{3, 2}$-3$_{2, 1}$ & 0.26(0.05) &   7.25(0.03) &   0.20(0.01) &               N   &         N   &         N   \\
c-C$_{3}$H$_{2}$ 6$_{0, 6}$-5$_{1, 5}$ &  0.20(0.04) &   7.20(0.02) &   0.21(0.02) &               N   &         N   &         N   \\
CH$_{3}$OH 1$_{0, 1}$-1$_{-1, 1}$     &            0.33(0.01) &              7.13(0.00) &              0.15(0.00) &        0.42(0.03) &  7.04(0.01) &  0.13(0.01) \\
CH$_{3}$OH 2$_{0, 2}$-2$_{-1, 2}$     &           0.12(0.01) &              7.13(0.01) &              0.13(0.01) &        0.21(0.03) &  7.05(0.02) &  0.13(0.02) \\
CH$_{3}$OH 3$_{0, 3}$-2$_{0, 2}$    &               0.08(0.01) &              7.18(0.01) &              0.15(0.01) &        0.14(0.03) &  7.12(0.02) &  0.10(0.02) \\
CH$_{3}$OH 3$_{0, 3}$-3$_{-1, 3}$     &                       $\lesssim$0.04 &                       - &                       - &                 $\lesssim$0.09&           - &           - \\
CH$_{3}$OH 3$_{1, 3}$-2$_{1, 2}$     &              0.78(0.01) &              7.17(0.00) &              0.15(0.00) &        1.06(0.02) &  7.10(0.00) &  0.12(0.00) \\
CH$_{3}$OH 4$_{0, 4}$-4$_{-1, 4}$     &                      $\lesssim$0.03 &                       - &                       - &                 $\lesssim$0.09 &           - &           - \\
HNCO 5$_{0, 5}$-4$_{0, 4}$           &                    0.60(0.04)   &                     7.18(0.02)   &                     0.18(0.02)   &                 0.85(0.02) &           7.16(0.02) & 0.17(0.02)       \\
HNCO 6$_{0, 6}$-5$_{0, 5}$           &           0.40(0.02)               &  7.18(0.02) &0.20(0.02) &        0.51(0.02) &  7.09(0.01) &  0.17(0.01) \\
HNCO 7$_{0, 7}$-6$_{0, 6}$           &               0.15(0.03) &              7.16(0.03) &              0.16(0.03) &        0.23(0.03) &  7.08(0.02) &  0.15(0.02) \\
HNCO 8$_{0, 8}$-7$_{0, 7}$           &               0.10(0.02) &              7.10(0.03) &              0.13(0.03) &                 $\lesssim$0.09 &           - &           - \\
\bottomrule
\end{tabular}
\begin{tablenotes}
\item Same as Table \ref{tab:gauss_lines}, continued.
\end{tablenotes}
\end{threeparttable}
\end{table*}

\subsubsection{RADEX models of CH$_{3}$OH, c-C$_{3}$H$_{2}$ and HNCO lines}\label{sec:radex}
We first use one-component non-LTE models to describe the emission of the observed lines.
Specifically, for HNCO, \ch\,and \choh\,molecules, we generated RADEX (\citealt{vdt07}) model grids, for column densities (N/$\Delta v$ with $\Delta v$ = 1 km s$^{-1}$) in the range of N = 10$^{11}$ to 10$^{15}\,$cm$^{-2}$ (with 60 logarithmically spaced uniform intervals), and hydrogen gas density in the range of 10$^{3}$ to 10$^{8}\,$cm$^{-3}$ (with 100 logarithmically spaced uniform intervals), and kinetic temperature in the range of 3$-$100 K (with 80 uniform intervals). The external radiation field was taken to be the cosmic background at 2.73 K.  The molecular data files for CH$_{3}$OH and HNCO were taken from the Leiden database, with collisional rates measured by \citet{RF} and \citet{hncocol}, respectively. For c-C$_{3}$H$_{2}$, we adopted the newly updated collisional rates which are down to 8 K (\citealt{Ben20}).

In the RADEX models, the line width is taken as 1 km/s for all the grids of parameters. The molecular column densities are further obtained by multiplying the fitted N/$\Delta v$ with the line width of each molecule. We used a linear interpolator to estimate the line intensities for parameters in between the intervals to better constrain the parameters and to allow for a continuous examination of parameter space.  
Nevertheless, the accuracy of the best-fit parameters remains limited by the resolution of the grid.  
We employ the Markov Chains Monte Carlo (MCMCs) method with an affine invariant sampling algorithm\footnote{A detailed description of the method can be found in the emcee documentation.} \cite{Foreman13} to perform the fitting (\citealt{Lin22}).
We initially kept all three parameters, $T_{\mathrm{kin}}$, $n(\mathrm{H_{2}})$ and $N_{\mathrm{mol}}$ as free parameters to fit with. For HNCO, we find the considered lines do not provide a good constraint on $T_{\mathrm{kin}}$. We then assume $T_{\mathrm{kin}}$ follows a normal distribution centering at 10 K and with a standard deviation of 5 K (N\,($\mu$=10 K, $\sigma$=5 K)). 
For the other parameters, the priors were assumed to be uniform distributions.
We use a likelihood distribution function which takes into account observational thresholds (detection limit, as listed in Tab. \ref{tab:gauss_lines}-\ref{tab:gauss_lines1}); the formulas follow
\begin{equation}
P\propto\underset{i}\Pi p_{i}\underset{j}\Pi p_{j}, 
\end{equation}
where $p_{i}$ stands for probabilities of the $i$th data that is a robust detection and $p_{j}$ the $j$th data that is an upper limit; we adopt the normal distribution as likelihood function,
\begin{equation}
p_{i} \propto \mathrm{exp}\,[-\frac{1}{2}(\frac{f^{\mathrm{obs}}_{i}-f^{\mathrm{model}}_{i}}{\sigma^{\mathrm{obs}}_{i}})^{2}]\Delta f_{i}
\end{equation}
\begin{equation}
p_{j} \propto \int _{-\infty}^{f^{\mathrm{obs}}_{lim,j}} \mathrm{exp}\,[-\frac{1}{2}(\frac{f_{j}-f^{\mathrm{model}}_{j}}{\sigma_{j}})^{2}]df_{j}, 
\end{equation}
in which $f^{\mathrm{obs}}_{i}$ (or $f_{j}$) stands for the observed intensity (or intensity upper limit) obtained from Gaussian fit, $f^{\mathrm{model}}_{i}$ (or $f^{\mathrm{model}}_{j}$) the model intensity, $\sigma^{\mathrm{obs}}_{i}$ is the standard error of the observed intensity which was adopted as the fitted 1$\sigma$ error of the Gaussian fit, $\sigma_{j}$ of the 1$\sigma$ noise level, $\Delta f_{i}$ being the data offset from the true value of $f_{i}$, and $d f_{j}$ the integrated flux probability to the detection threshold $f^{\mathrm{obs}}_{lim,j}$.

The starting points (initialization) for the chains were chosen to be the parameter set corresponding to a global $\chi^{2}$ minimum calculated between the grid models and the observed values.  
The ``burn-in" phase in the MCMC chains is included and we also conduct several resets of the starting-points to ensure the final chains are reasonably stable around the maximum of the probability. The posterior distributions of the parameters for all the molecules and at different locations are shown in Figure \ref{fig:corners}-\ref{fig:corners3}. The best-fit parameters are listed in Table \ref{tab:radex_para}, for the dust peak and the molecular peaks. 

To account for the different beam sizes of the observed lines and the possible beam dilution effect, we have tested the modeling by excluding the coarser resolution lines in Table \ref{tab:lines_more}, specifically, for CH$_{3}$OH the three $J$=2-1 lines at $\sim$ 96 GHz, and for c-C$_{3}$H$_{2}$, the 3$_{1,2}$-2$_{2,1}$ line. We find that the derived parameters are similar to the fiducial case where the lines are included in the fit (without correction of beam dilution). The emission of these lower frequency lines are likely more extended than the respective finest beam size listed in Table \ref{tab:lines_more}, as also indicated by the 30m mapping observations in \citet{Spezzano17} and the NOEMA maps in \citet{Punanova18} such that the beam dilution effect is not significant. For HNCO lines, if we assume the emitting area is comparable to the finest beam among the lines ($\sim$14$''$), and scale the intensities of lines with coarser resolution, we arrive at lower gas densities ($\lesssim$10$^{4}$ cm$^{-3}$). However, again, these lower frequency lines likely have extended emission that make the beam dilution effect not as significant as the results derived by the re-scaled intensities. Therefore in Table \ref{tab:radex_para} we list derived parameters based on line intensities without beam dilution correction. In Sect. \ref{sec:loc}, we use full radiative transfer models and generate line cubes following the structure of L1544, and account for the varying beam sizes for comparisons with observations (Appendix B).

\begin{figure}
\hspace{-.75cm}\includegraphics[scale=0.265]{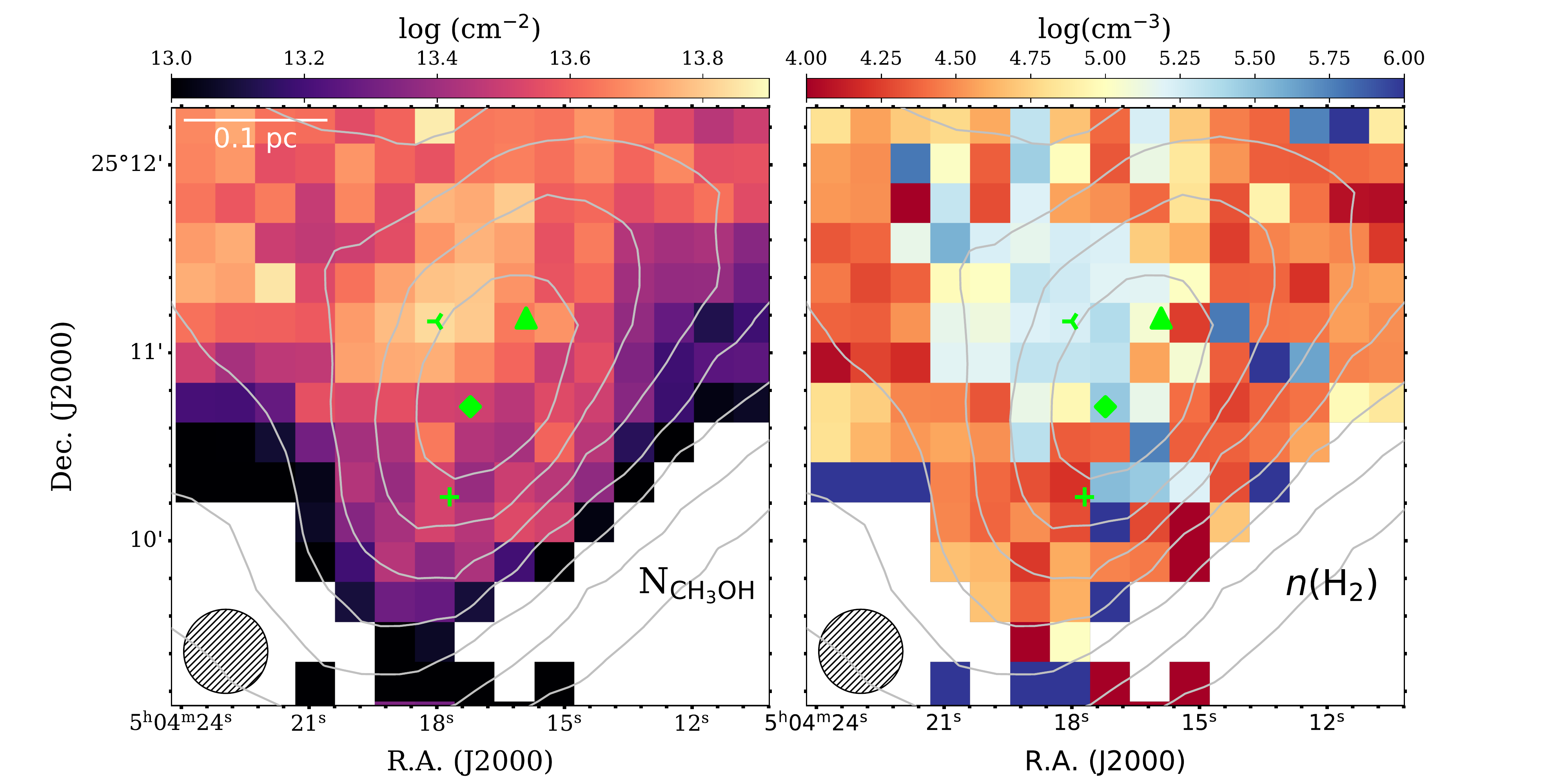} 
\caption{Derived CH$_{3}$OH column density map and hydrogen volume density from RADEX modeling. The contours (in gray) are $N_{\mathrm{H_{2}}}$ levels,  same as that in Fig. \ref{fig:guidemaps}. The markers in blue represent the dust peak (diamond), molecular emission peaks of CH$_{3}$OH (three-branched triangle),  HNCO (upper triangle) and c-C$_{3}$H$_{2}$ (plus).}
    \label{fig:n_nmol_map}
\end{figure}

In addition, the mapping observations of $J$\,=\,2-1 CH$_{3}$OH $E$ type lines, in which 2$_{-1,2}$-1$_{-1,1}$ and 2$_{0,2}$-2$_{0,1}$ were detected, are used to derive an $n(\mathrm{H_{2}})$ map and $N_{\mathrm{CH_{3}OH}}$ map. These maps are shown in Fig. \ref{fig:n_nmol_map}. It can be seen from the $n(\mathrm{H_{2}})$ map that there is a prominent density enhancement centralised at the CH$_{3}$OH peak. The regions with enhanced gas densities at the dust peak, HNCO and c-C$_{3}$H$_{2}$ peaks are less extended than toward the CH$_{3}$OH peak. The $N_{\mathrm{CH_{3}OH}}$ values in Fig. \ref{fig:n_nmol_map} are in the range of 1.0-7.9$\times$10$^{13}$ cm$^{-2}$, which are consistent with that reported by \citet{Bizzocchi14} and \citet{Vastel14}.

\begin{table*}
\centering
\scriptsize
\begin{threeparttable}
\caption{Derived gas properties using RADEX models at the dust peak and molecular peaks.}
\label{tab:radex_para}
\begin{tabular}{llllllllll}
\toprule
&\multicolumn{9}{c}{Molecule}\\
&\multicolumn{3}{c}{\choh}&\multicolumn{3}{c}{c-\ch}&\multicolumn{3}{c}{HNCO}\\
\\
Position&$n(\mathrm{H_{2}})$&$T_{\mathrm{kin}}$&$N_{\mathrm{mol}}$&$n(\mathrm{H_{2}})$&$T_{\mathrm{kin}}$&$N_{\mathrm{mol}}^{a}$&$n(\mathrm{H_{2}})$&$T_{\mathrm{kin}}$&$N_{\mathrm{mol}}$\\
&cm$^{-3}$&K&cm$^{-2}$&cm$^{-3}$&K&cm$^{-2}$&cm$^{-3}$&K&cm$^{-2}$\\
\midrule
Dust peak&4.7$\times$10$^{5}$(0.05)&7.4(0.2)&1.3$\times$10$^{13}$(0.05)&5.9$\times$10$^{4}$(0.3)&13.2(1.2)&1.4$\times$10$^{13}$(0.2)&6.9$\times$10$^{4}$(0.4)&11.8(1.0)$^{b}$&4.3$\times$10$^{12}$(0.1)\\
\choh\, peak&3.4$\times$10$^{5}$(0.05)&16.5(0.2)&1.6$\times$10$^{13}$(0.03)&\\ 
c-\ch\, peak&6.7$\times$10$^{5}$(0.2)&8.4(0.7)&0.9$\times$10$^{13}$(0.07)&7.0$\times$10$^{3}$(0.3)&15.2(0.7)$^{b}$&1.5$\times$10$^{14}$(0.3)&\\
HNCO peak&6.3$\times$10$^{5}$(0.1)&15.0(1.5)&1.2$\times$10$^{13}$(0.03)&&&&4.4$\times$10$^{4}$(1.1)&11.6(3.0)$^{b}$&8.2$\times$10$^{12}$(0.1)\\
  \bottomrule
\end{tabular}
    \begin{tablenotes}
      \small
      \item  $^{a}$: The column densities of c-C$_{3}$H$_{2}$ are listed for the total column density, assuming an ortho- to para-c-C$_{3}$H$_{2}$ ratio of 3;  $^{b}$: $T_{\mathrm{kin}}$ follows a prior distribution of N\,($\mu$=10 K, $\sigma$=5 K).     
      \item  Values in parentheses are estimated 1$\sigma$ errors from the posterior distribution function. For $n(\mathrm{H_{2}})$ and $N_{\mathrm{mol}}$ which were sampled in log scale, 1$\sigma$ errors are given in proportion, with respect to the parameter values. 
      \end{tablenotes}
        \end{threeparttable}
\end{table*}

\begin{figure*}
    \centering
    \hspace{-0.5cm}\includegraphics[scale=0.35]{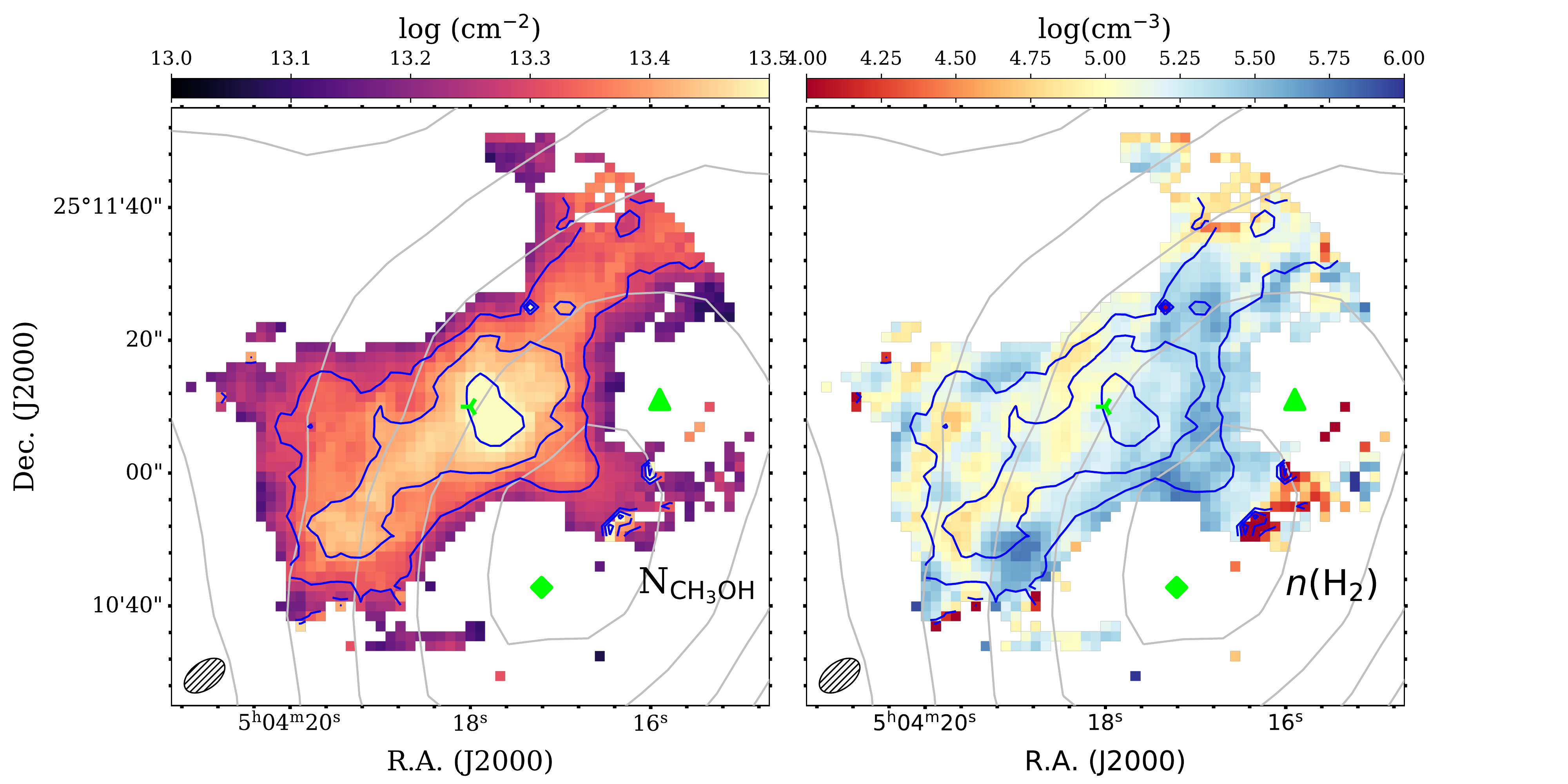}
    \caption{The $N_{\mathrm{CH_{3}OH}}$ and $n(\mathrm{H_{2}})$ maps at the CH$_{3}$OH peak (data from \citealt{Punanova18}). The blue contour levels correspond to $N_{\mathrm{CH_{3}OH}}$ values of 2.0$\times$10$^{13}$,  2.5$\times$10$^{13}$ and 3.2$\times$10$^{13}$ cm$^{-2}$. The gray contours show the $N(\mathrm{H_{2}})$ contours of 1.0$\times$10$^{22}$, 1.2$\times$10$^{22}$, 1.6$\times$10$^{22}$, 2.0$\times$10$^{22}$, 2.5$\times$10$^{22}$ cm$^{-2}$, from the SED fitting of {\it{Herschel}} data (\citealt{Spezzano16}). The dust peak, CH$_{3}$OH peak and the HNCO peak are indicated as green diamond, three-branched triangle and upper triangle, respectively.}
    \label{fig:noema_n_Nmol}
\end{figure*}

To further probe the physical properties at the CH$_{3}$OH peak in detail, we used the high angular resolution observations of CH$_{3}$OH $J$\,=\,2-1 lines towards the CH$_{3}$OH peak (\citealt{Punanova18}). The spectral cube contains data combined from NOEMA and 30m telescope, achieving a resolution of $\sim$700 au. The derived $n(\mathrm{H_{2}})$ map and $N_{\mathrm{CH_{3}OH}}$ maps are shown in Fig. \ref{fig:noema_n_Nmol}. The density enhancements in the $n(\mathrm{H_{2}})$ map are located mostly in the south-west direction of the column density peak, facing the core center and reaching $\sim$10$^{6}$ cm$^{-3}$. 

\subsubsection{Full radiative transfer modeling with LOC}\label{sec:loc}
While with RADEX modeling we constrain physical parameters with a one-component non-LTE radiative transfer approximation that retains the assumption of local excitation, to understand the mutual impact of molecular abundance profiles and gas density distribution on the resultant line intensities (and ratios), we adopt full radiative transfer calculations using LOC (line transfer with OpenCL, \citealt{Juvela20}). In LOC, deterministic ray tracing and accelerated lambda iterations (\citealt{RH91}) are employed for modeling the radiative transfer process. We use the 1D model in LOC which considers spherically symmetric distribution of physical structures, characterised by volume density, $\rho(r)$, kinetic temperature, $T(r)$ and velocity field, including both micro-turbulence, $\sigma_{\mathrm{turb}}$ and radial velocity, $V(r)$. 

\begin{figure}
    \centering
    \includegraphics[scale=0.5]{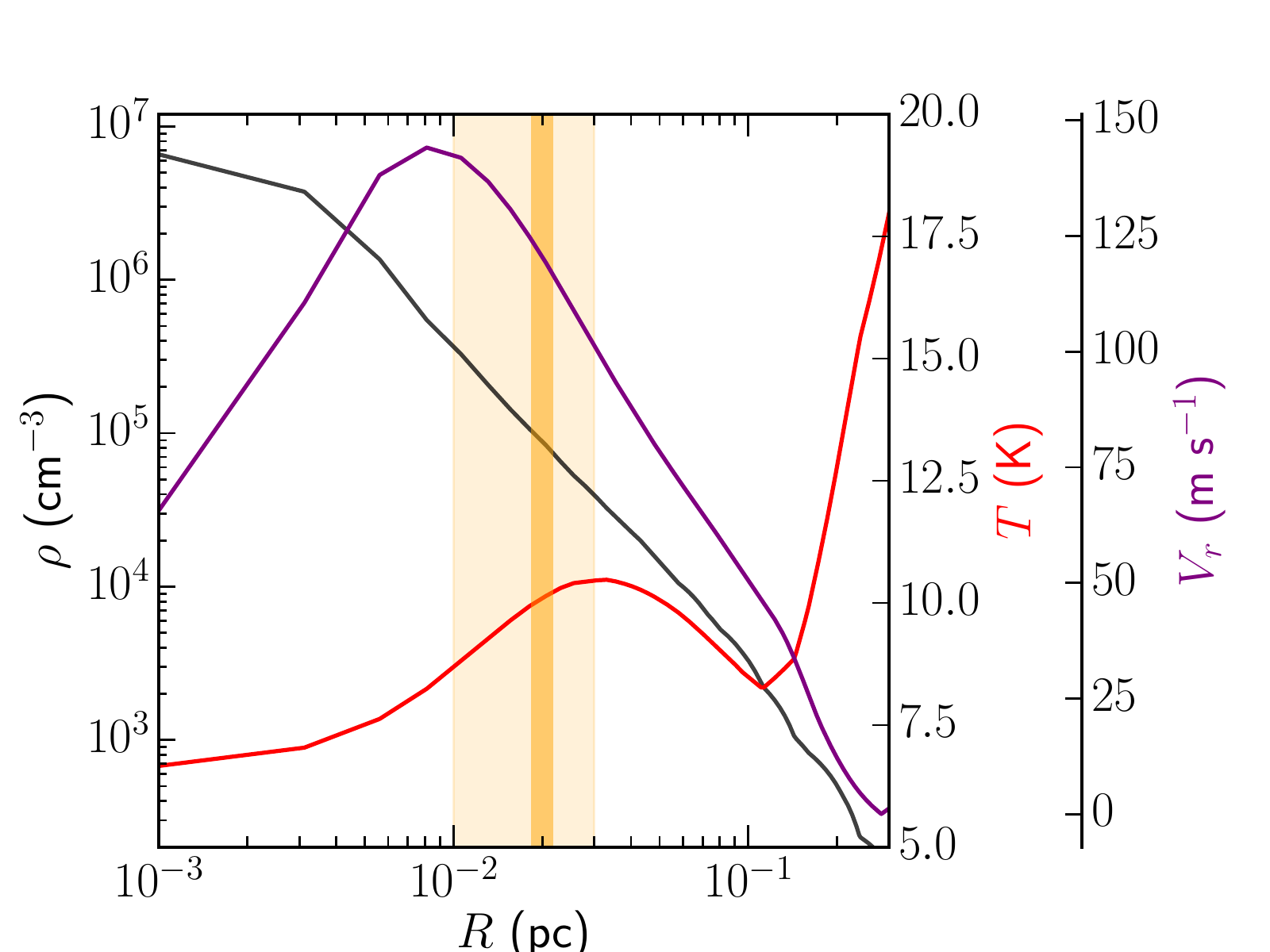}
    \caption{The radial profiles of gas density, temperature and infall velocity. The light orange shaded region indicates the beam area with which the molecular peaks are identified, centered at an offset of 0.02 pc (with respect to the dust peak), shown by a vertical orange line.}
    \label{fig:rho_T_v}
\end{figure}

\begin{figure*}
    \centering
    \hspace{-0.15cm}\includegraphics[scale=0.425]{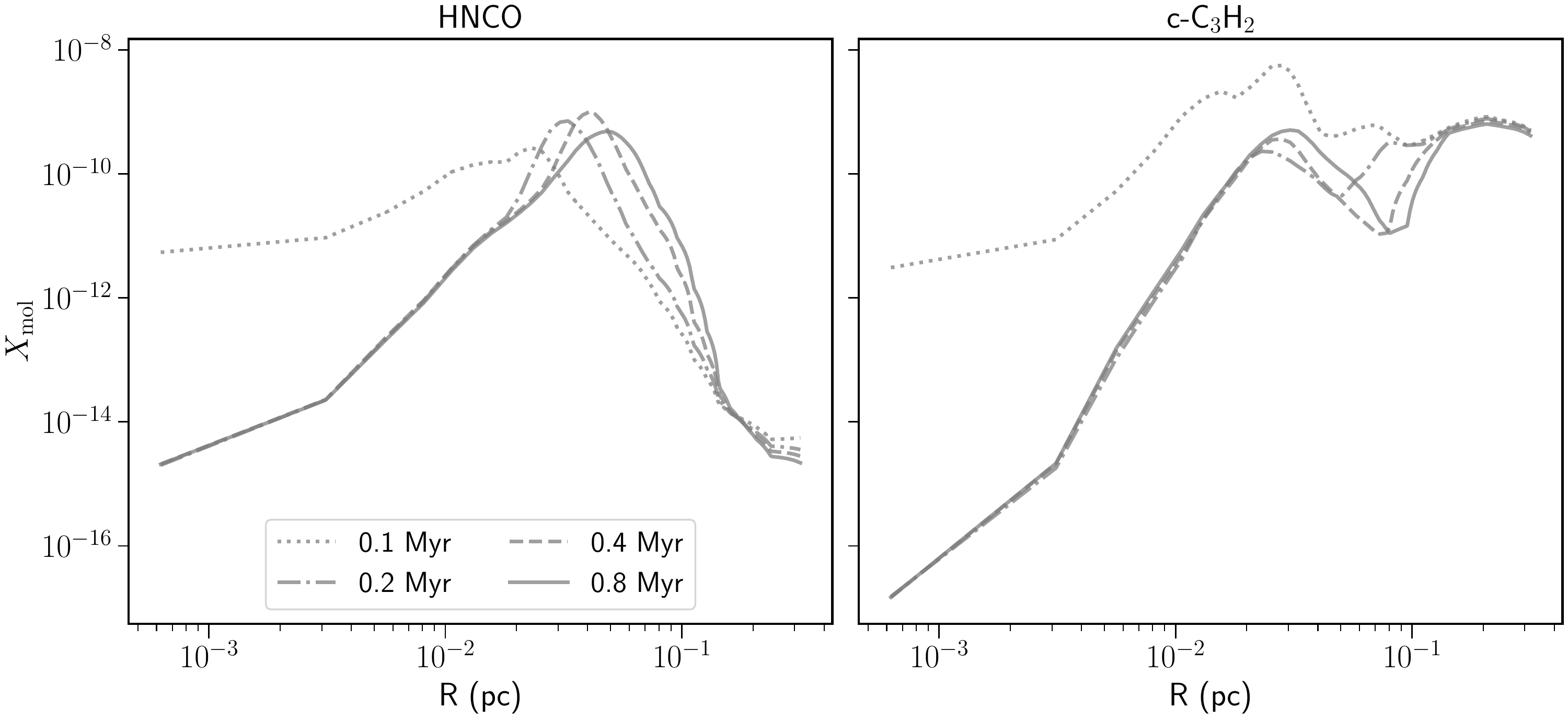}\\
        \hspace{-0.15cm}\includegraphics[scale=0.425]{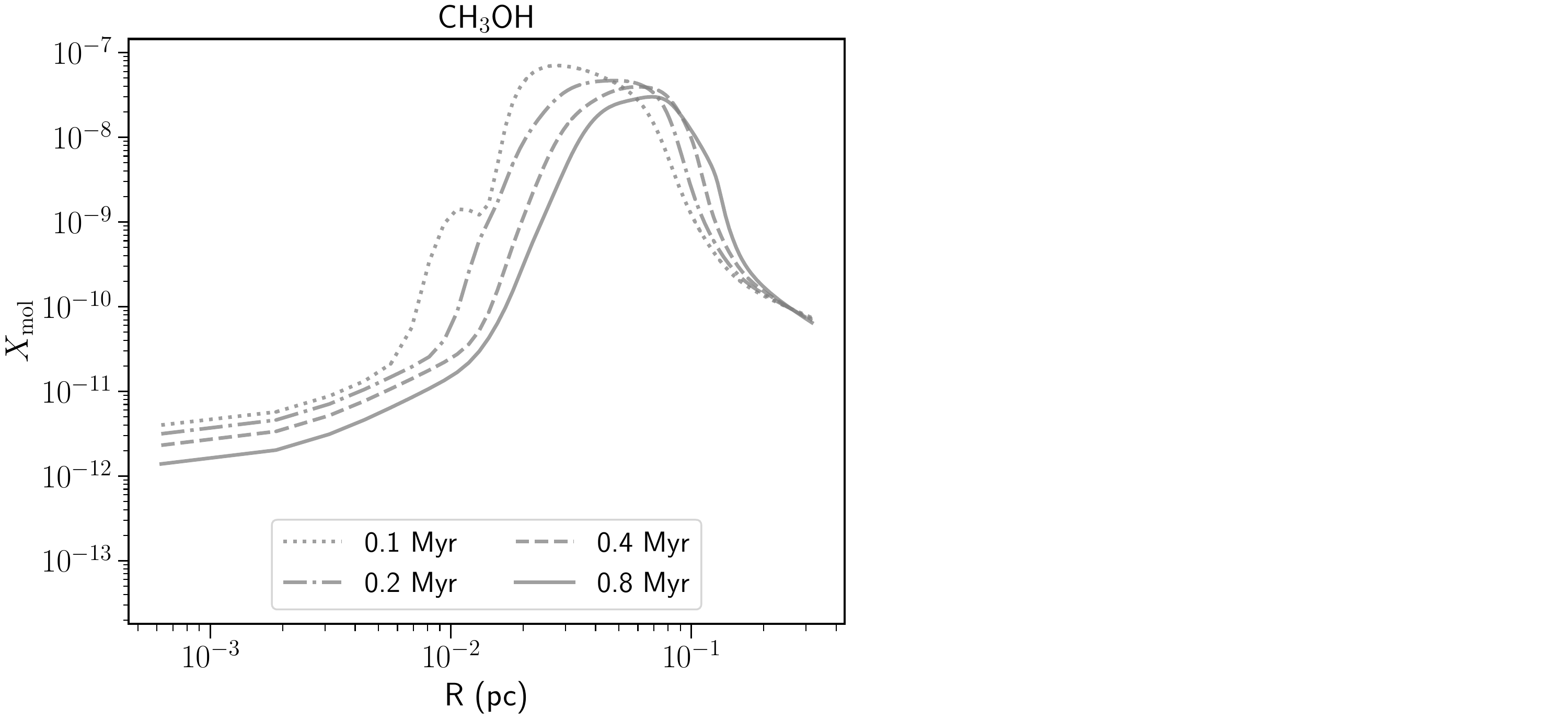}
    \caption{{\emph{Upper panel:}} The radial abundance profiles of c-C$_{3}$H$_{2}$ and HNCO for selected epochs, predicted by chemical modeling in \citet{Sipila15}. {\emph{Lower panel:}} The radial abundance profiles of CH$_{3}$OH, predicted by chemical modeling in \citet{vasyunin17}.}
    \label{fig:ab_prof_select}
\end{figure*}

We adopted the parameterised forms of $\rho(r)$, $T(r)$ and $V(r)$ following \citet{KC10} for L1544, which are shown in Fig. \ref{fig:rho_T_v}. 
The core radius is assumed to be 0.3 pc. In the modeling, linear discretisation is used for the grids with a physical resolution of 60 au. We then convolved the output spectral cube with the corresponding observational beam for each frequency. 
We started the modeling with the radial abundance profiles predicted from the chemical models in \citet{vasyunin17} for CH$_{3}$OH and from \citet{Sipila15, Sipila16} for \ch\ and HNCO. Examples of the abundance profiles are shown in Fig. \ref{fig:ab_prof_select}, for epochs of 0.1, 0.2, 0.4, and 0.8 Myr.  For CH3OH abundance profiles, we adopted the results from \citet{vasyunin17} due to the fact that the treatment of CH3OH formation and reactive desorption is more comprehensive as they took into account the change in reactive desorption with different surface compositions following experimental work by \citet{Minissale16b}. In both models, a static physical model of the core is adopted following \citet{KC10} and the chemistry is treated independently for different gas density layers. Initial elemental abundances are based on \citet{WH08} (see also \citealt{Semenov10}).  For more details of the two chemical models, as well as the differences between them, we refer the readers to the original papers of \citet{Sipila15, Sipila16} and \citet{vasyunin17}.

\begin{figure*}[htb]
\begin{tabular}{p{0.45\linewidth}p{0.45\linewidth}}
\includegraphics[scale=0.5]{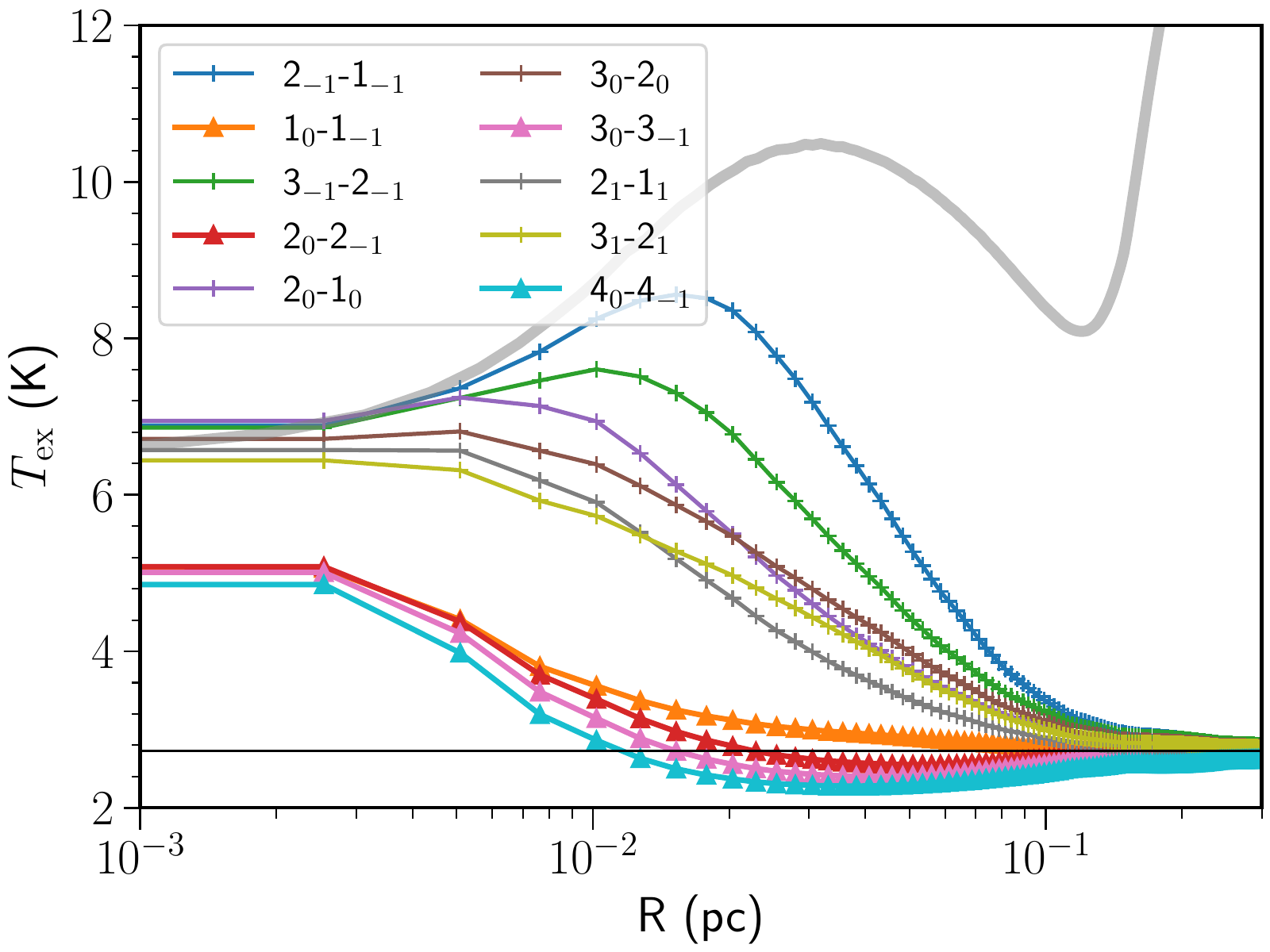}& \includegraphics[scale=0.5]{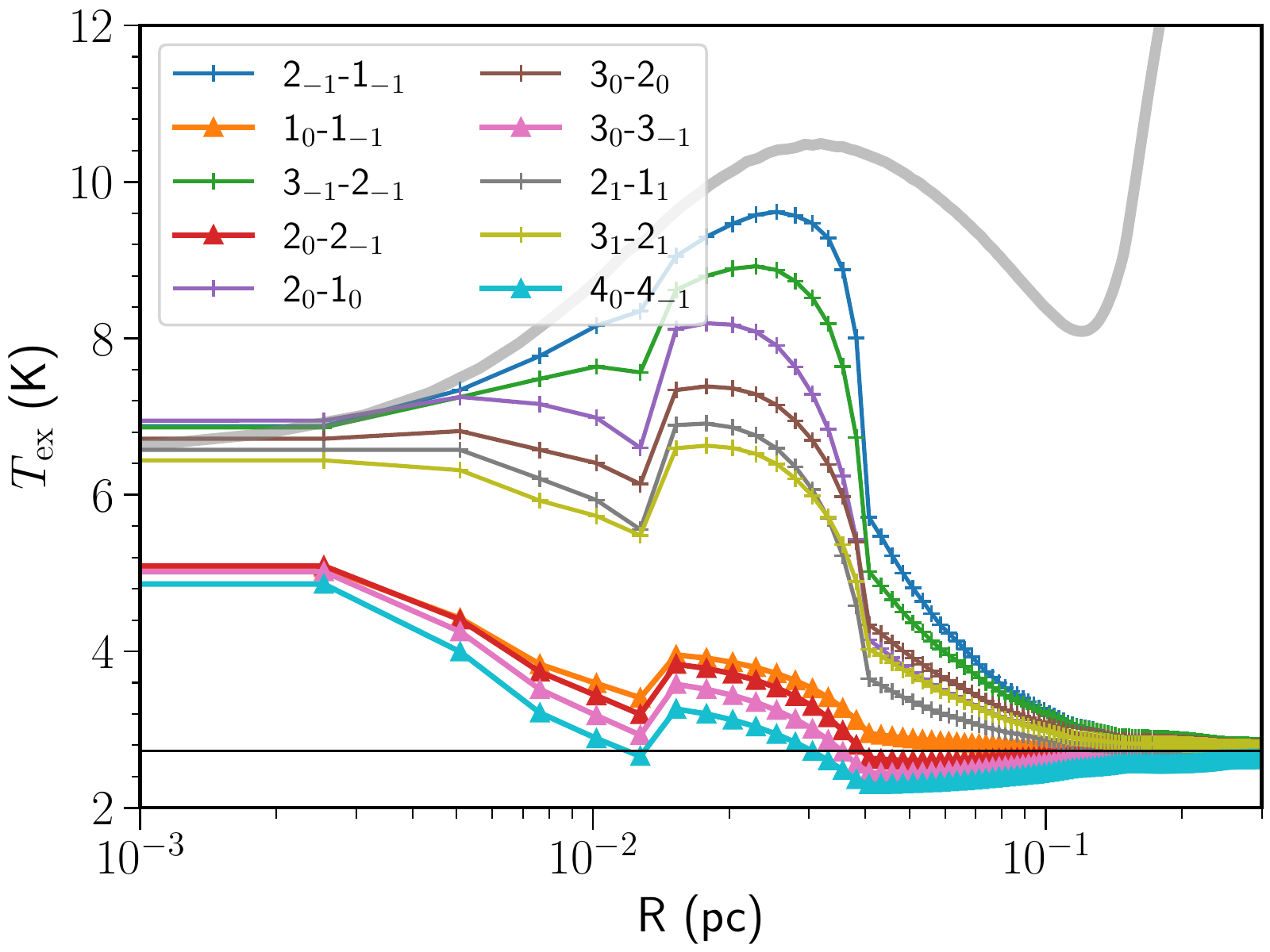}\\
\end{tabular}
    \caption{The comparison between $T(r)$ (gray line) and radial profile of $T_{\mathrm{ex}}$ of all CH$_{3}$OH lines from the best-fit model with modified radial density profile (left figure) and original density profile (right). The best-fit abundance profiles are at 0.8 Myr from \citet{vasyunin17} which were scaled down by a factor of 5 and 10 in the left and right panel (of model A and model B), respectively. The CMB temperature of 2.73 K is also indicated (black horizontal line). The four $J_{0}$-$J_{-1}$ lines are shown with thicker line widths.}
    \label{fig:tex_bump}
\end{figure*}

Comparisons between the modelled spectra and the observed line intensity are shown in Fig. \ref{fig:vas_ch3oh_model}-\ref{fig:olli_hnco_model_adj} in Appendix B. For CH$_{3}$OH, in general, the abundance profiles from chemical models produce spectra in excess of intensities and show significant self-absorption, which are not reflected by the observations. The four to five times larger column densities predicted by the chemical models compared to that estimated from observations of CH$_{3}$OH in L1544 (\citealt{Vastel14}) was also noted by \citet{vasyunin17} initially. However, merely scaling down the abundance profile does not give a much better match (see Fig. \ref{fig:vas_ch3oh_model_adjab} where the abundance is scaled down by a factor of 5, here after model A). In particular, the intensities of CH$_{3}$OH 3$_{0}$-3$_{-1}$ and 2$_{0}$-2$_{-1}$ around $V_{\mathrm{LSR}}$ still show negative emission. Examination of the excitation temperature profiles shows that for these two lines $T_{\mathrm{ex}}$ drops below the cosmic microwave background temperature ($T_{\mathrm{CMB}}$) beyond $\sim$0.015 pc (Fig. \ref{fig:tex_bump}, left panel), becoming anti-inverted. Also, the intensity of 1$_{0}$-1$_{-1}$ line is significantly underestimated compared to the other lines.  Given the characteristics of the energy levels of $E$-type CH$_{3}$OH (\citealt{Lees73}, \citealt{Kalenski16}), this is somehow expected, since these transitions have upper levels in the side ladders and lower levels in the main ladder, and the latter tends to be over-populated, causing a low $T_{\mathrm{ex}}$ of the transitions when the gas densities are low.  

\begin{figure}[htb]
\hspace{1cm}\includegraphics[scale=0.45]{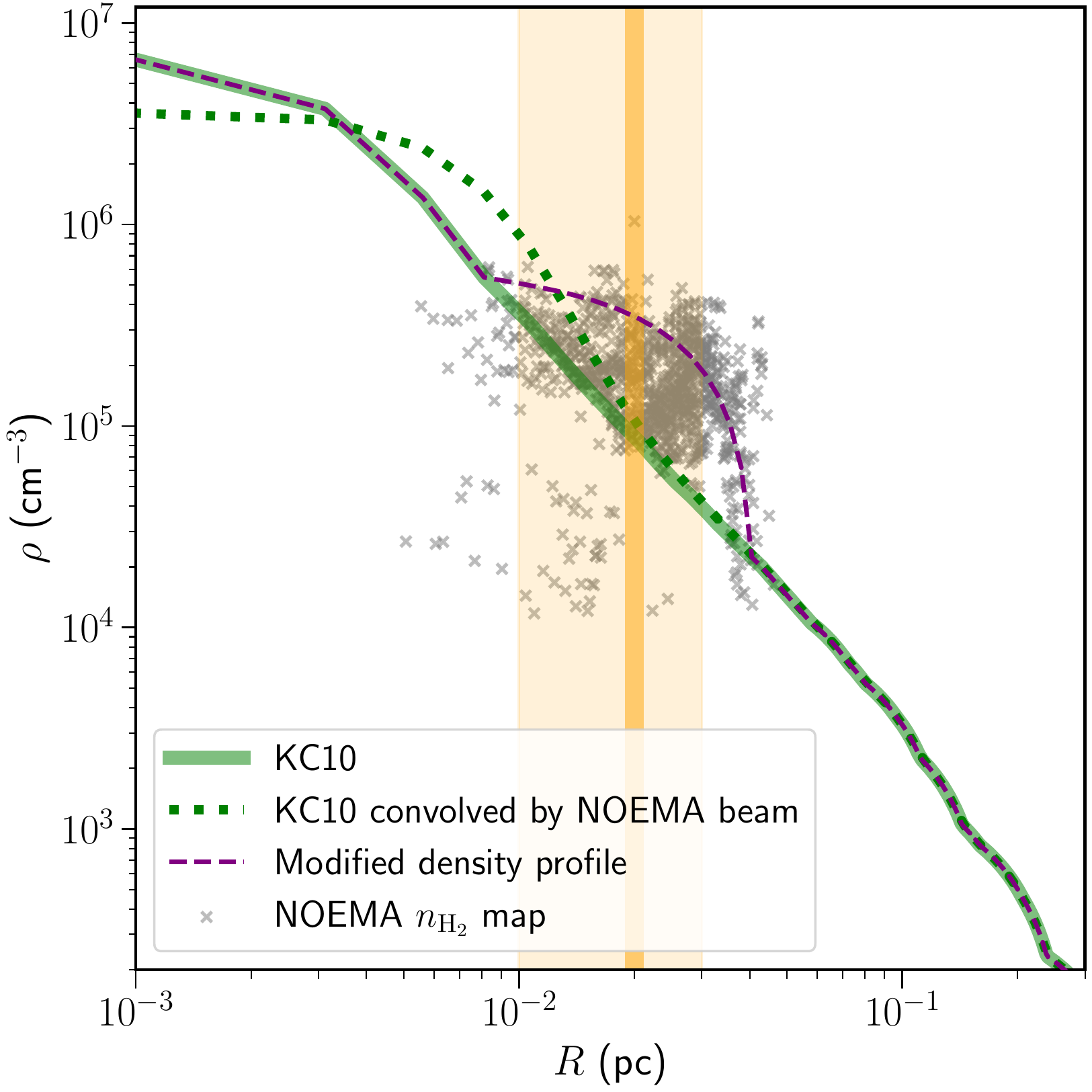}
\caption{Modified radial density profile (purple) in comparison with the original and beam convolved radial density profiles (green) of L1544 established in \citet{KC10} (indicated by KC10 in the legend). Gray crosses indicate the gas density values measured from the $n(\mathrm{H_{2}})$ map as in Fig. \ref{fig:noema_n_Nmol}.  The light orange shaded region indicates the beam area with which the molecular peaks are identified, centered at an offset of 0.02 pc (with respect to the dust peak), shown by a vertical orange line.}
\label{fig:dens_bump}
\end{figure}
\begin{figure*}
    \centering
    \includegraphics[scale=0.33]{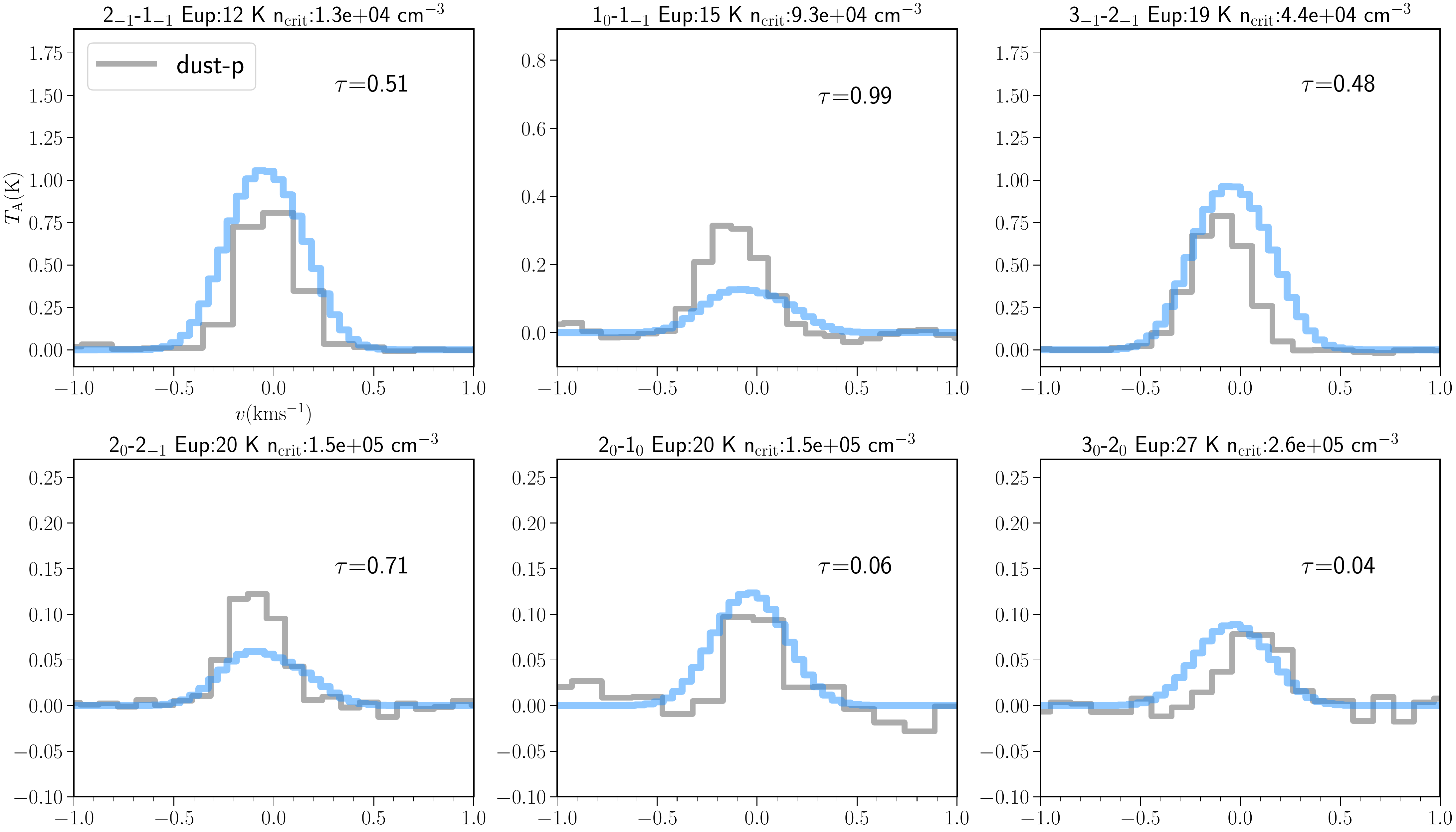}\\
    \includegraphics[scale=0.33]{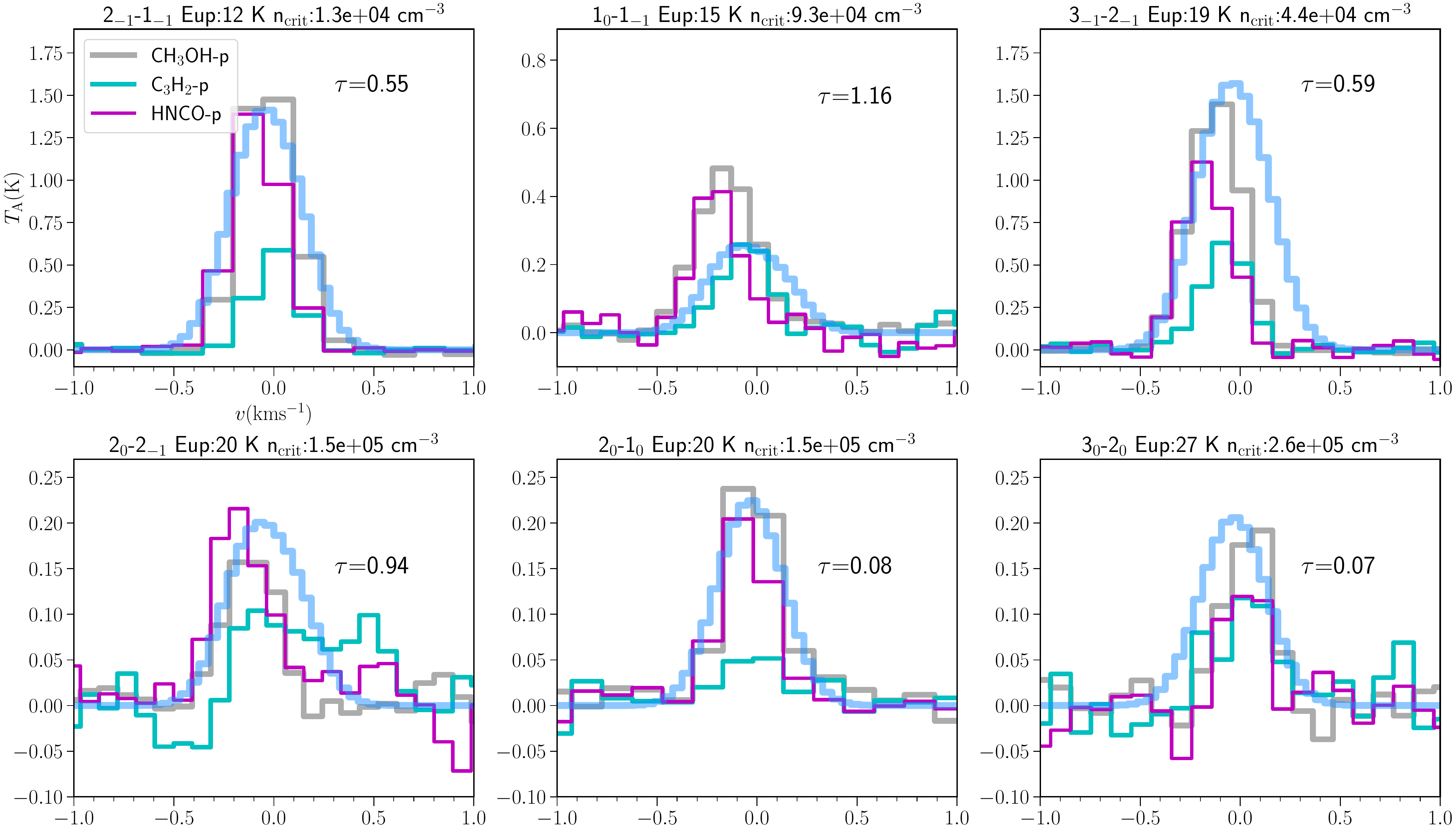}
    
    \caption{Comparison between the best-fit modelled spectra (in blue lines) and observed CH$_{3}$OH lines (gray histograms) at the dust peak (upper two panels) and CH$_{3}$OH peak (lower two panels). In the lower panel, the spectra at the c-C$_{3}$H$_{2}$ peak and HNCO peak are also shown for comparison. The optical depth at the line center derived from the model is indicated in each subplot.}
    \label{fig:ch3oh_6best}
\end{figure*}

The uncertainties of the abundance of CH$_{3}$OH of current chemical models are mainly due to the assumptions on reactive desorption rates as well as on the adoption of a static core  (\citealt{vasyunin17}). 
As noted by \citet{vasyunin17}, when dynamical evolution is taken into consideration, during which the gas density is enhanced due to gravitational collapse, there will be less CH$_{3}$OH produced on grains at early times, since CO freeze-out is less efficient in a low-density environment.
Noting this and the differences between modelled spectra and observations, we attempted to adjust the combination of radial abundance profile and gas density profile to generate spectra that match with the observed intensities. Based on the RADEX results, the $n(\mathrm{H_{2}})$ seen by CH$_{3}$OH lines at all molecular peaks are similar with that at the dust peak. It is likely that at these molecular peaks, which are $\sim$0.02 pc from the dust peak, there are localised gas density enhancements causing the emission of the CH$_{3}$OH lines (see also \citealt{Bacmann16} for CH$_{3}$OH lines in other pre-stellar cores).  We therefore modified the radial gas density profile (hereafter modified density model) by adding a local density increment of 5$\times$10$^{5}$ cm$^{-3}$ with a width of 0.025 pc centered at $\sim$0.03 pc from the dust peak. The width is chosen to match with the $n(\mathrm{H_{2}})$ map from NOEMA observations (Fig. \ref{fig:dens_bump}), and is close to the 30m beam HPFW at the frequencies of 2$_{K}$-1$_{K}$ lines. The resultant `perturbed' radial density profile is shown in Fig. \ref{fig:dens_bump}.  Such a modified density profile, in combination with a scaled-down (by a factor of 10) radial abundance profile (at 0.8 Myr epoch from \citealt{vasyunin17}), can reproduce the intensities of most CH$_{3}$OH lines well (hereafter model B), and the 1$_{0}$-1$_{-1}$ and 2$_{0}$-2$_{-1}$ lines within a factor of $\sim$2. With the enhanced density, the 2$_{0}$-2$_{-1}$ line now turns into emission and the 1$_{0}$-1$_{-1}$ line exhibits higher intensity (Fig. \ref{fig:ch3oh_6best}). We note that the parameters used for the modified density model is not a unique combination: a change in the width and/or absolute value of the density enhancement with a different abundance profile may also produce spectra that fit well with the observed lines. The point is that 
varying only the abundance profiles we cannot qualitatively match with the CH$_{3}$OH emission lines.
Our choice of parameters is guided by the $n(\mathrm{H_{2}})$ map derived from the high-angular resolution NOEMA observations. In \citealt{vasyunin17}, formation of CH$_{3}$OH in gas phase is most efficient at gas densities of $\sim$10$^{4}$-10$^{5}$ cm$^{-3}$ where CO molecules start to catastrophically freeze out. Indeed, from the NOEMA results of the $n(\mathrm{H_{2}})$ and $N_{\mathrm{CH_{3}OH}}$ maps (Fig. \ref{fig:noema_n_Nmol}), the highest $n(\mathrm{H_{2}})$ values do not correspond to the highest $N_{\mathrm{CH_{3}OH}}$. This is a consistent result showing that CH$_{3}$OH is likely more enhanced in moderate gas density regimes, while an enhanced gas density with a possible local reduction of gas temperature due to increased gas-dust coupling can promote the depletion of CH$_{3}$OH. The impact of the density enhancement on the physical and chemical properties of L1544 should be investigated in future with high angular resolution observations.

\begin{figure}[htb]
    \includegraphics[scale=0.5]{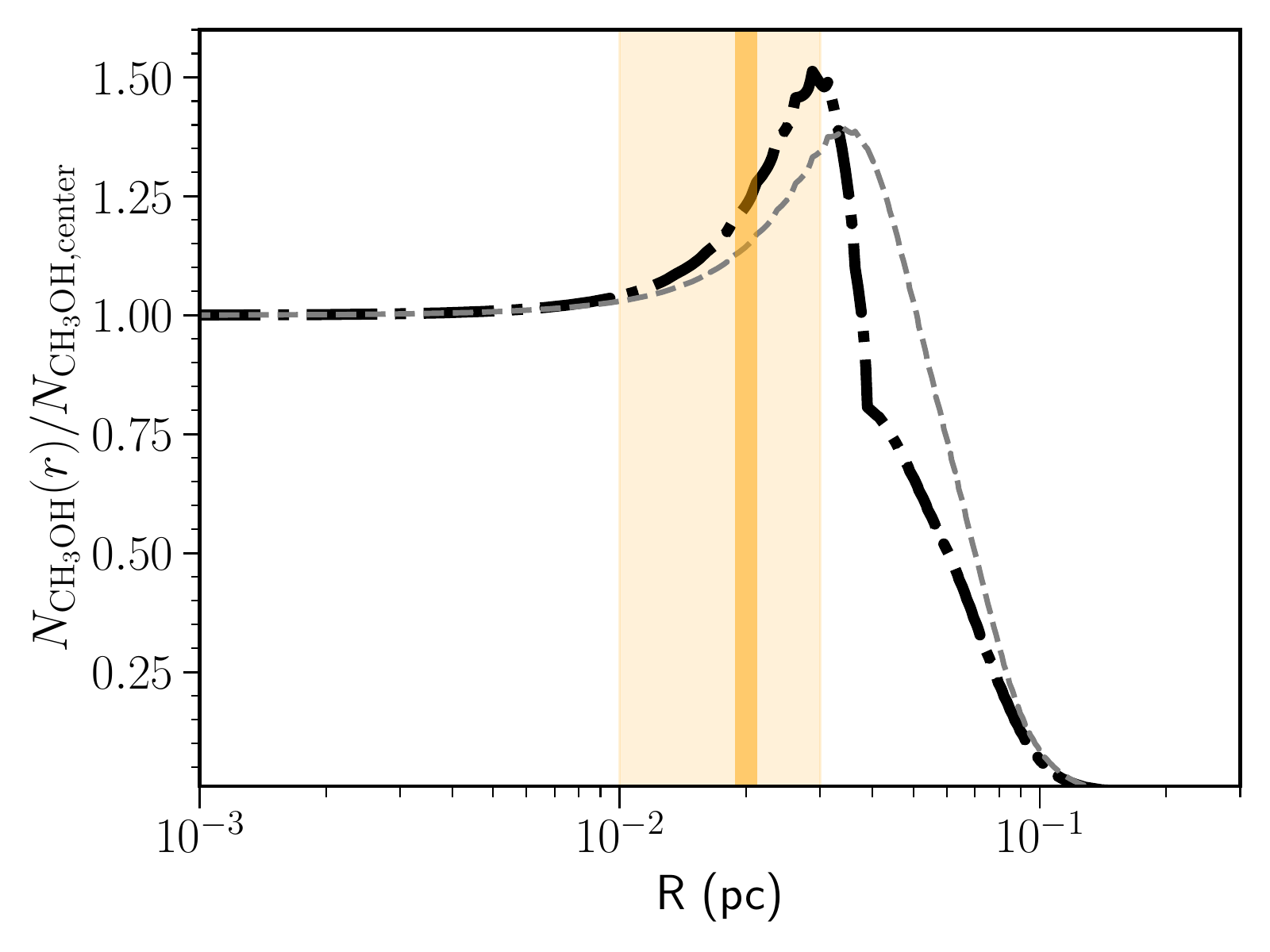}
    \caption{The comparison between radial profile of CH$_{3}$OH column density, $N_{\mathrm{CH_{3}OH}}$ normalised by central value, $N_{\mathrm{CH_{3}OH,center}}$ from the best-fit model with original radial density profile (model A, gray dashed line) and modified radial density profile (model B, black dash-dotted line). The orange shaded region and vertical line follow same definition as in Fig. \ref{fig:rho_T_v}}
    \label{fig:nmol_bump}
\end{figure}

The inclusion of a density enhancement that better matches with the data is mainly driven by the comparisons with the 2$_{0}$-1$_{-1}$ and 1$_{0}$-1$_{-1}$ lines, which have the highest $n_{\mathrm{crit}}$ of all CH$_{3}$OH lines. The four $J_{0}$-$J_{-1}$ lines are the most sub-thermally excited lines, as can be seen from their $T_{\mathrm{ex}}$ profile in comparison with $T(r)$ in Fig. \ref{fig:tex_bump}. We compared the $T_{\mathrm{ex}}$ profiles from the modified density profile model and the original density model in Fig. \ref{fig:tex_bump}. It can be seen that the inclusion of the density enhancement at 0.015-0.040 pc leads to an increase of $T_{\mathrm{ex}}$ at these radii, which levels up the emission of the strongly sub-thermally excited lines to match with the observations. The modelled radial profiles of $N_{\mathrm{CH_{3}OH}}$ from best-fit model A and model B are shown in Fig. \ref{fig:nmol_bump}: with the density enhance, the peak of $N_{\mathrm{CH_{3}OH}}$ is also more prominent and appears closer to the 0.02 pc radius.


For c-C$_{3}$H$_{2}$ and HNCO, the abundance profiles from chemical models in \citet{Sipila15} can reproduce the observed line intensities relatively well (Fig. \ref{fig:olli_c3h2_model} and Fig. \ref{fig:olli_hnco_model}). The line intensities at the dust peak can be well predicted within 30$\%$ by the abundance profile at 0.1 Myr, while the line intensities at the HNCO peak are systematically underestimated within a factor of two. We find that a factor of 2-5 increment of abundance profile will match with the observations at HNCO peak better, depending on the epoch of the abundance profile used.  
The results are shown in Fig. \ref{fig:olli_c3h2_model_adj}. 
For the c-C$_{3}$H$_{2}$ lines, the modelled spectra show the best consistency with the observed line intensities (Fig. \ref{fig:olli_c3h2_model}), similar with what was presented by \citealt{Sipila16}. The observed line intensities at the dust peak are within $\sim$20$\%$ of the modelled spectra for the abundance profile at 0.1 Myr. For the c-C$_{3}$H$_{2}$ peak, a factor of three increment of abundance profile can produce the observed line intensities better, which is shown in Fig. \ref{fig:olli_c3h2_model_adj}.  The fact that HNCO and c-C$_{3}$H$_{2}$ lines do not need a modification of density profile to fit with the model spectra indicates that these lines indeed mostly originate from the bulk volume of gas in the outer regions of L1544, without a prominent preferential emission region of either denser nor diffuse gas layers.  This corroborates our analysis from the RADEX modeling, described in Sec. \ref{sec:radex}. 


\section{Discussion}\label{sec:dis}
\subsection{CH$_{3}$OH, c-C$_{3}$H$_{2}$ and HNCO: tracing gas layers of different densities across L1544}

From the radiative transfer modeling, we find that the gas densities seen by CH$_{3}$OH, c-C$_{3}$H$_{2}$ and HNCO are very different (Table \ref{tab:radex_para}). The lines of CH$_{3}$OH are excited at highest gas densities, reaching up to 3-7$\times$10$^{5}$ cm$^{-3}$, both at the dust peak and at the molecular peaks. Such high density gas at the dust peak and c-C$_{3}$H$_{2}$ peak is also associated with a low kinetic temperature below 10 K, while at the HNCO and CH$_{3}$OH peaks the kinetic temperature is $\sim$15 K. On the other hand, modeling of the c-C$_{3}$H$_{2}$ and HNCO lines indicate gas densities of several times 10$^{4}$ cm$^{-3}$, suggesting that these lines mainly originate from the outer, less dense layers of the pre-stellar core. Towards dust peak, the kinetic temperatures indicated by the HNCO and c-C$_{3}$H$_{2}$ lines, although not well constrained, are likely above 10 K (Fig. \ref{fig:corners1}-\ref{fig:corners3}). This is consistent with the fact that the outer gas layers of pre-stellar cores are of higher temperatures with respect to the central region (with temperatures close to 6 K; \citealt{Crapsi07}). 

The C$_{3}$H$_{2}$ lines have been suggested to be good density probes for dark molecular clouds (10$^{4}$-10$^{5}$ cm$^{-3}$, \citealt{Avery87}, \citealt{Cox89}). The gas densities towards starless and protostellar cores probed by multiple transitions of c-C$_{3}$H$_{2}$ are usually several 10$^{4}$ cm$^{-3}$ (\citealt{Cox89}, \citealt{Takakuwa01}). Our results are consistent with these previous findings. The full radiative transfer modeling of LOC also suggests that the two lowest energy transitions of c-C$_{3}$H$_{2}$ are heavily optically thick ($\tau$=0.8-1.2, Fig. \ref{fig:olli_c3h2_model}-\ref{fig:olli_c3h2_model_adj}), at both the dust peak and c-C$_{3}$H$_{2}$ peak, which indicates they mostly originate from the outer parts of the core, tracing the sub-thermally excited gas ($n_{\mathrm{crit}}$ in Table \ref{tab:lines_more}). 
The lack of correlation between c-C$_{3}$H$_{2}$ deuterium fraction and the central H$_{2}$ density in a sample of pre-stellar cores (\citealt{Chantzos18}) is also an evidence of the origin of c-C$_{3}$H$_{2}$ emission in outer layers.

The emission of CH$_{3}$OH lines in dense cores is widely observed (e.g., \citealt{Tafalla06}, \citealt{Soma15}, \citealt{Vastel14}, \citealt{Bizzocchi14}, \citealt{Harju20}, \citealt{Spezzano20}, \citealt{Punanova21}). In starless core TMC-1 CP, CH$_{3}$OH lines indicate a gas density of several times 10$^{4}$ cm$^{-3}$, which is consistent or lower than that probed by carbon-chain molecules (\citealt{Soma15}). While TMC-1 CP is prior to gravitational collapse and likely in an earlier evolutionary stage than L1544, as indicated by its chemical composition (e.g. \citealt{Pratap97}, \citealt{Agundez19}, \citealt{Navarro21}), in L1544 the $\sim$10 times higher gas densities seen by CH$_{3}$OH might be the result of the evolution of the core. \citet{Bacmann16} derived similar gas densities with CH$_{3}$OH lines towards the centers of a sample of pre-stellar cores. Indeed, by introducing the concept of ``contribution function'', \citealt{Tafalla06} show that emission of the CH$_{3}$OH lines in non-LTE condition mainly arises from the high density gas layers of pre-stellar cores. For L1544, the radial gas density profile constrained by full radiative transfer models of both dust and multiple molecular lines in \citealt{KC10}, when convolved with the 30m beam at the frequency of the 2$_{k}$-1$_{k}$ lines, gives a gas density of 5.0$\times$10$^{5}$ cm$^{-3}$ at the three molecular peaks, and 8.4$\times$10$^{5}$ cm$^{-3}$ at the dust peak. These values are consistent with gas densities seen by CH$_{3}$OH at these positions. This means that despite the central depletion of CH$_{3}$OH, the line emission is still dominated by the dense gas proportion inside the beam, i.e. the inner gas layer close to the CH$_{3}$OH depletion zone. We discuss the local gas density enhancement seen by CH$_{3}$OH in Sec. \ref{sec:dens_enhance}.


Although the three molecular species indicate distinct gas volume densities, the lines do not exhibit difference in line widths nor centroid velocities. This is likely due to both the relatively low velocity resolution we have ($\sim$0.1 km s$^{-1}$) and the fact that contraction velocities of L1544 are subsonic with a velocity peak at 1000 au (\citealt{KC10}), such that the single-dish observations of the three molecules are not sensitive to gas kinematics. Nevertheless, with the combination of multiple lines and radiative transfer calculations, we obtained the tomographic view towards L1544, in which the three molecular species are ``peeling off" different outer gas density layers by their different weighting schemes of emission contribution along the core radii. 
Also, the results demonstrate that single-dish observations with molecular line emission, in particular, certain transitions of CH$_{3}$OH (e.g., the $J_{0}$-$(J-1)_{-1}$ lines), enable us to pinpoint the small-scale enhanced gas density distribution inside pre-stellar cores, which is not readily seen in the previous Herschel column density map and/or ground-based continuum observations.
The full RT modeling is important to disentangle the effect of varying radial abundance from gas densities in determining the line excitation characteristics.

\subsection{Local density enhancement at methanol peak}\label{sec:dens_enhance}

The one-component non-LTE models of CH$_{3}$OH lines at the three molecular peaks suggest gas density well above 10$^{5}$ cm$^{-3}$. With LOC models, we find that only by increasing the gas densities locally can the observed line intensities and ratios be well reproduced. This was already suggested by the result of RADEX models: even with the depletion of CH$_{3}$OH in the core center such that the line emission does not sample the densest gas, beam averaged gas densities (from the one-component non-LTE models) are still comparable to that estimated from the radial density profile of L1544 without incomplete sampling. Previous works on a variety of molecular lines and dust continuum towards L1544 have consistently suggested that the {\emph{overall}} density structure of the core follows a BE sphere (\citealt{KC10}, \citealt{Keto15}, \citealt{R19}, \citealt{K20}).  The density enhancement seen by CH$_{3}$OH suggests that the density structures can deviate from the average BE density profile {\emph{locally}}, showing a clumpier distribution. 

\citet{Punanova18} mapped the CH$_{3}$OH peak with the NOEMA observations and found that the centroid velocity and velocity dispersion increase towards the dust peak indicative of accretion or interaction of two filaments. The large-scale cloud L1544 embedded in has two perpendicular structures (\citealt{Andre10}, \citealt{Spezzano16}) which seem to meet at the CH$_{3}$OH peak (see also \citealt{Tafalla98}), and may cause a weak collision shock. In Sec. \ref{sec:radex}, using the NOEMA observations, we obtained the $n(\mathrm{H_{2}})$ map using the ratio of the 2$_{0}$-1$_{0}$ and 2$_{-1}$-1$_{-1}$ lines of $E$ CH$_{3}$OH. Within the uncertainties, the $n(\mathrm{H_{2}})$ map shows that the regions of largest densities form a ring-like structure surrounding the central core region of L1544, which do not coincide with the $N_{\mathrm{CH_{3}OH}}$ peak (Fig. \ref{fig:noema_n_Nmol}). Overall, the morphology of the density enhancements around the CH$_{3}$OH peak of L1544 appears rather distributed and spatially extended, without clear and concentrated centers. 
Since the density enhancements are also associated with largest velocity in the south-west direction ($\sim$7.3 km s$^{-1}$, see the Fig. 6 of \citealt{Punanova18}), the picture seems to favor the mild accretion flow which accelerates towards the direction of the core center, causing enhanced gas densities. At this position, the merging of large-scale cloud structures may also play a role in delivering the gas material, which remains to be investigated with extended mapping to reveal the kinematic features.

The density enhancement in the modified density models is a factor of 5-10 larger than the densities at same radii predicted from BE sphere model. We now consider whether it could be a result of shock conditions.
In the presence of shocks (either due to accretion or merging clouds), the density enhancement across an interface is related to variations of fluid velocity (\citealt{Draine93}), where $\rho_{0}v_{0}$ = $\rho_{1}v_{1}$. With an additional condition of energy conservation in the case of adiabatic shocks, it holds that $\rho_{1}$/$\rho_{0}$ = $\frac{(\gamma+1)M_{0}^{2}}{(\gamma+1)+(\gamma-1)(M_{0}^{2}-1)}$ in which $\gamma$ is the adiabatic index and $M_{0}$ the Mach number. The maximum density contrast for diatomic gas ($\gamma$ = 7/5) is therefore 6. On the other hand, for isothermal shocks,  $\rho_{1}$/$\rho_{0}$ = $M_{0}^{2}$, meaning that the density enhancement can be arbitrarily large. Taking the temperature of 10 K at 0.02 pc with a sound speed of 0.2 km s$^{-1}$, for a density enhancement of a factor of 5-10, the Mach number should be $\gtrsim$4.5 in the adiabatic case and $\sim$2.2-3 in the isothermal case. The velocity gradient surrounding the CH$_{3}$OH peak appears largest in the south of the map (close to the dust peak), reaching $\sim$12 km s$^{-1}$ pc$^{-1}$ (\citealt{Punanova18}). Over a distance of 0.025 pc, this corresponds to a velocity difference of $\sim$0.3 km s$^{-1}$ assuming it varies uniformly. Considering projection effect, a real velocity variation of $\sim$0.6-1 km s$^{-1}$ is possible, reaching the Mach number required for the density enhancement we resolved here.

Towards the HNCO and c-C$_{3}$H$_{2}$ peaks, the density enhancement is less spatially extended than that at the CH$_{3}$OH peak. The density enhancements at 0.02 pc thus appear asymmetric, based on measurements on these three directions of molecular peaks. A high angular resolution and sensitivity mapping of CH$_{3}$OH lines is desired given the preferential distribution of CH$_{3}$OH, since at the HNCO and c-C$_{3}$H$_{2}$ peak the higher $K$ components of CH$_{3}$OH line might be too weak to yield more robust determinations of $n(\mathrm{H_{2}})$. 

The localised density enhancement likely only occupies a small volume of the core, such that the overall density structure is well described by a BE sphere. The non-uniform distribution of dense gas surrounding the central region of dense cores may be related to the highly variable accretion rates associated with the formation of low-mass protostars at smaller scales, a common phenomenon during the protostar evolution (\citealt{Audard14}).  Based on the association of the density enhancement with the gas velocity gradient, we speculate the clumpy gas can be formed in situ due to nonsteady gravitational inflow, and it exists upon the overall smooth BE sphere of the bulk gas. 
Numerical simulations also show that clumpy gas around an early-stage core can originate initially due to disturbance of turbulence (\citealt{Matsumoto11}, \citealt{Lewis18}), which tend to be smoothed out inside a gravitationally collapsing core. Whether the over-densities will be enhanced due to self-gravity or torn apart by tidal forces depends, essentially, on the their mass scale and ``thickness'' with respect to that of the enclosed gas (i.e. gas component interior to the over-density, \citealt{Coughlin17}). The eventual infall of the over-densities onto the central point mass can cause a temporal increase of the accretion rate. 
At any rate, a smoothly decreasing BE sphere is not sufficient to describe the localised gas behaviour inside the pre-stellar core as well as the influence from large scale cloud environment.  How the underlying over-densities of a pre-stellar core form initially, and how they evolve and eventually affect star formation remain to be systematically investigated in future observational and theoretical studies.

\section{Conclusions}\label{sec:conclusion}
 
Using multi-line observations of CH$_{3}$OH, c-C$_{3}$H$_{2}$ and HNCO towards the dust peak and the three respective molecular peaks of L1544, different gas density layers are probed. With full radiative transfer modeling implemented with radial abundance profiles, we revisit the density structure of L1544 and gauge the current chemical models. The main conclusions are: 
\begin{enumerate}
    \item With one-component non-LTE models, we find that at the dust peak and the three molecular peaks at 0.02 pc radius, CH$_{3}$OH lines are tracing gas densities of $n(\mathrm{H_{2}})$$>$10$^{5}$ cm$^{-3}$, while emission lines of c-C$_{3}$H$_{2}$ and HNCO mainly originate from the outer, less dense gas layers of several 10$^{4}$ cm$^{-3}$. Even with the incomplete sampling of the inner core region due to CH$_{3}$OH depletion, the $n(\mathrm{H_{2}})$ traced by CH$_{3}$OH lines are close to the beam-averaged gas densities at the dust peak and molecular peaks estimated from the previously well-established radial density profiles of L1544. 
    \item The BE sphere density structure of L1544, coupled with the radial abundance profiles of HNCO and c-C$_{3}$H$_{2}$ predicted from chemical models (\citealt{Sipila16}), can produce spectra consistent with the observed line intensities at the dust peak and respective molecular peaks. The abundance profiles only require a factor of 3-5 increment to achieve better agreement with the observed lines at the molecular peaks, within the uncertainty of the chemical models.
    \item At the CH$_{3}$OH peak, the gas density enhancements seen from the $n(\mathrm{H_{2}})$ map derived from the CH$_{3}$OH 2$_{K}$-1$_{K}$ lines at $\sim$700 au angular resolution are offset from the $N_{\mathrm{CH_{3}OH}}$ peak, and appear to be a ring-like clumpy structure surrounding the core center. 
    \item CH$_{3}$OH lines at the dust peak and CH$_{3}$OH peak can only be well reproduced with a modified radial density model that includes a local density enhancement encompassing the $\sim$0.02 pc radius of the core in addition with a scaled-down (by a factor of 5-10) radial abundance profile of CH$_{3}$OH based on chemical models (\citealt{vasyunin17}). The existence of such a local density enhancement may be caused by slow shocks induced by asymmetric and dynamic accretion flows, and may be related to the kinematics of the large-scale cloud where L1544 is embedded. 
\end{enumerate}

In this work, we demonstrated that multi-line observations of molecules that exhibit significant chemical differentiation can be adopted as useful tools to probe the underlying physical properties in pre-stellar cores. CH$_{3}$OH lines, in particular, may be well suited to probe the dense gas component of pre-stellar cores. Our results again suggest that a symmetric and smooth density profile can be over-simplified to describe the structure of pre-stellar cores, which may exhibit over-densities locally that deviate from the overall BE sphere structure. A more extended and higher angular resolution CH$_{3}$OH mapping towards L1544 is desirable to pinpoint the origin of the over-densities and shed light on the physical and chemical processes causing the enhancement of CH$_{3}$OH in gas phase, which could be related to the on-going accretion flows onto the pre-stellar core.






     

\begin{acknowledgements}
      The authors acknowledge the financial support of the Max Planck Society. 
      Y. Lin thanks the helpful discussion with Mika Juvela.
      This work is based on observations carried out under project number 101-20 with the IRAM 30m telescope. IRAM is supported by INSU/CNRS (France), MPG (Germany) and IGN (Spain).\end{acknowledgements}

\bibliography{ref}

\begin{appendix}
\section{Posterior distribution of the MCMC RADEX models}
\begin{figure*}
\begin{tabular}{p{0.45\linewidth}p{0.45\linewidth}}
    \includegraphics[scale=0.4]{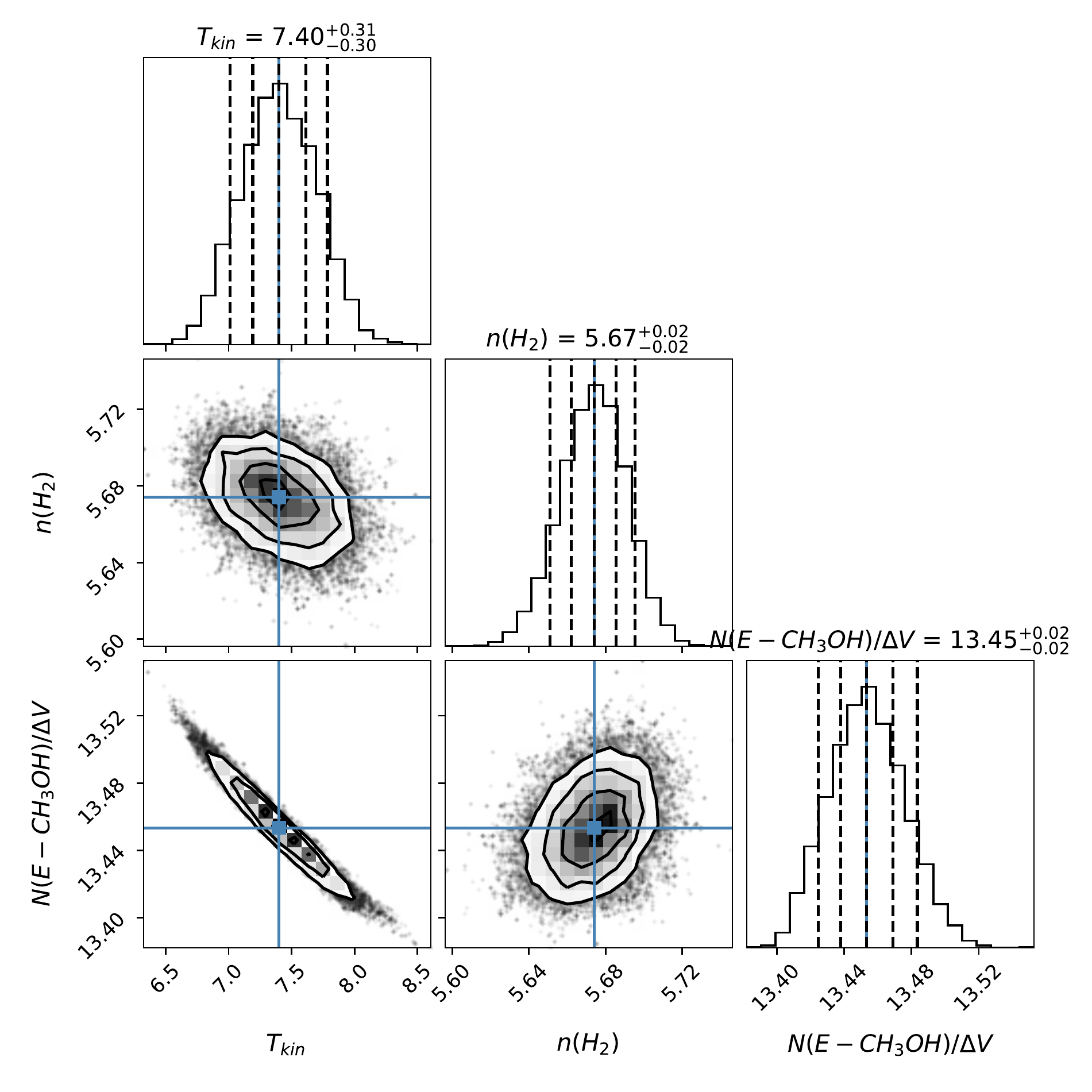}&\includegraphics[scale=0.4]{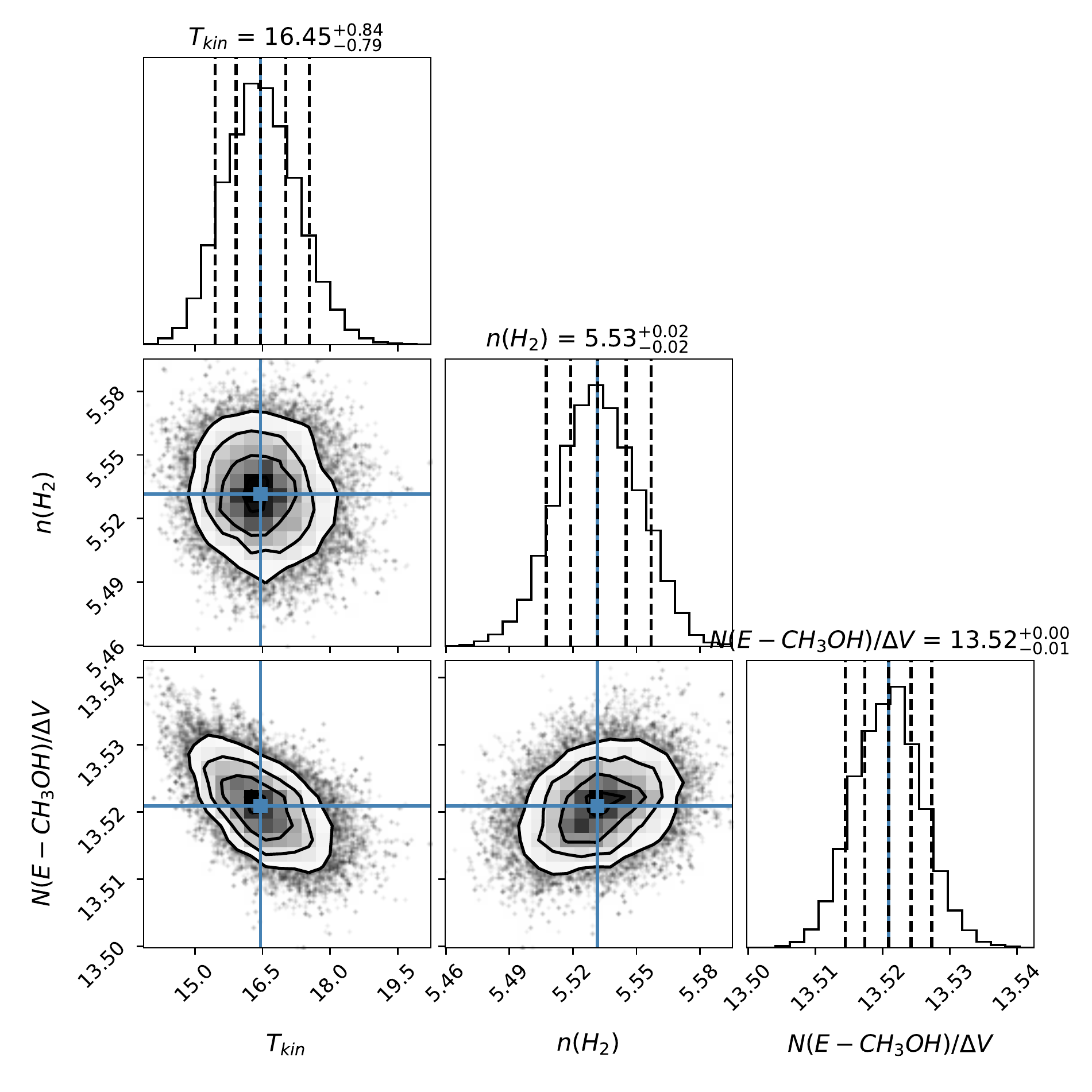}\\
    \includegraphics[scale=0.4]{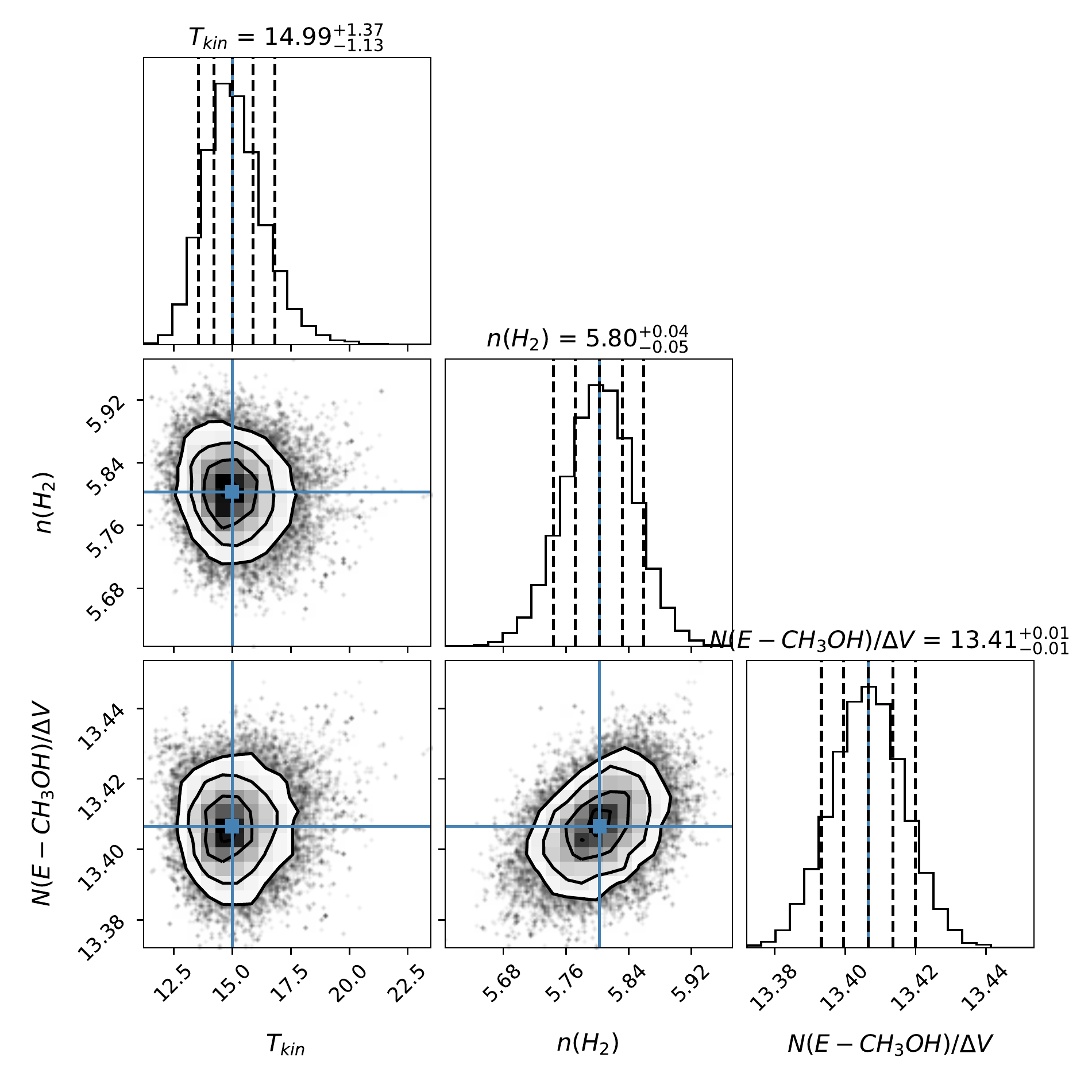}&\includegraphics[scale=0.4]{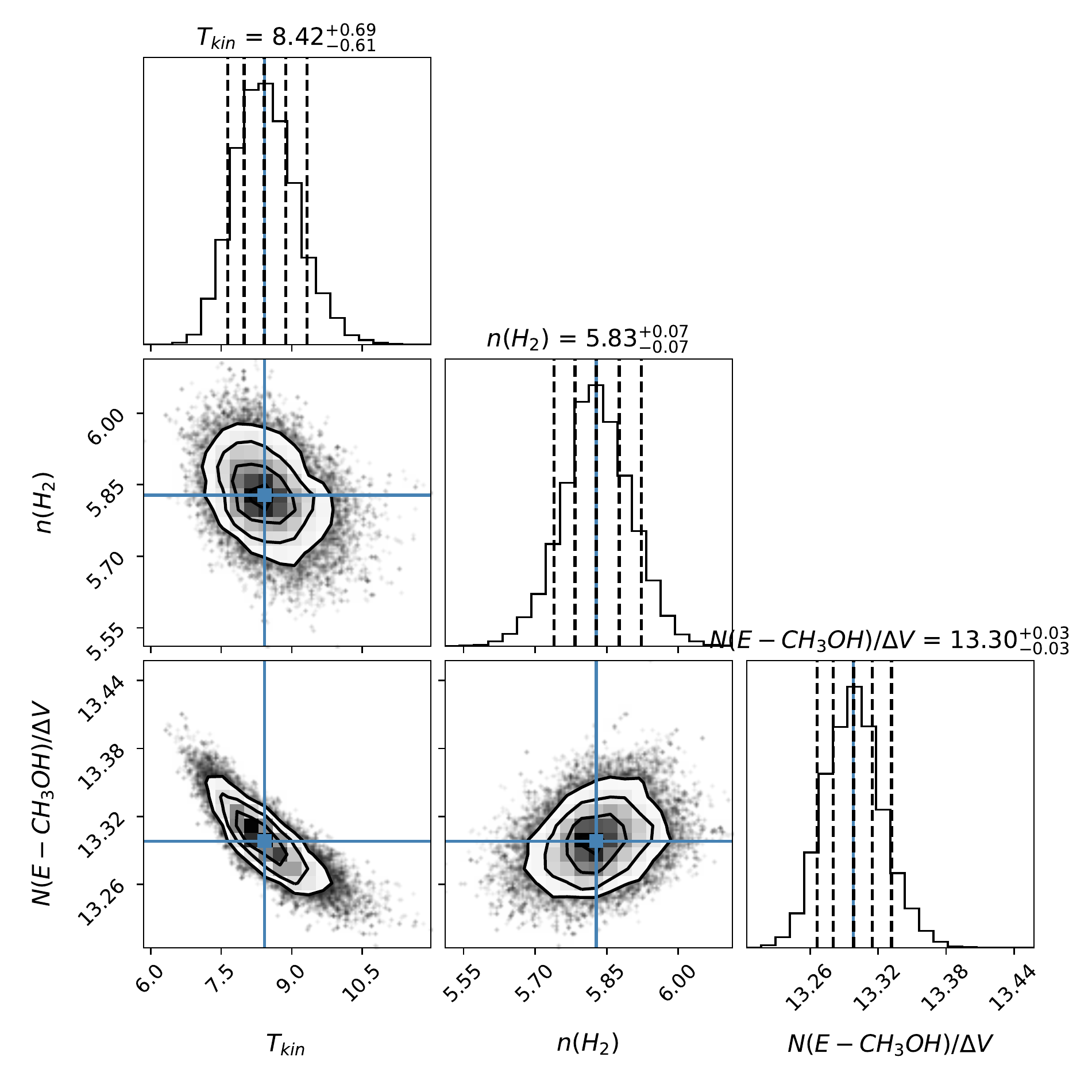}\\
    \end{tabular}
    \caption{Posterior distribution of parameters $T_{\mathrm{kin}}$, log$_{10}$($n(\mathrm{H_{2}})$), log$_{10}$($N_{\mathrm{mol}}$/$\Delta V$) based on modeling CH$_{3}$OH ($E$-type) lines at the dust peak (upper left), CH$_{3}$OH peak (upper right), HNCO peak (lower left) and c-C$_{3}$H$_{2}$ peak (lower right). All three parameters are kept free. In these plots and plots in Figs. \ref{fig:corners1}-\ref{fig:corners3}, the vertical dashed lines in the 1d histograms show the quantiles of 10$\%$, 25$\%$, 50$\%$, 75$\%$, 90$\%$. The contour levels in the 2D histograms indicate 0.5$\sigma$, 1$\sigma$, 1.5$\sigma$ and 2$\sigma$, respectively.}
    \label{fig:corners}
\end{figure*}

\begin{figure*}
\begin{tabular}{p{0.45\linewidth}p{0.45\linewidth}}
    \includegraphics[scale=0.4]{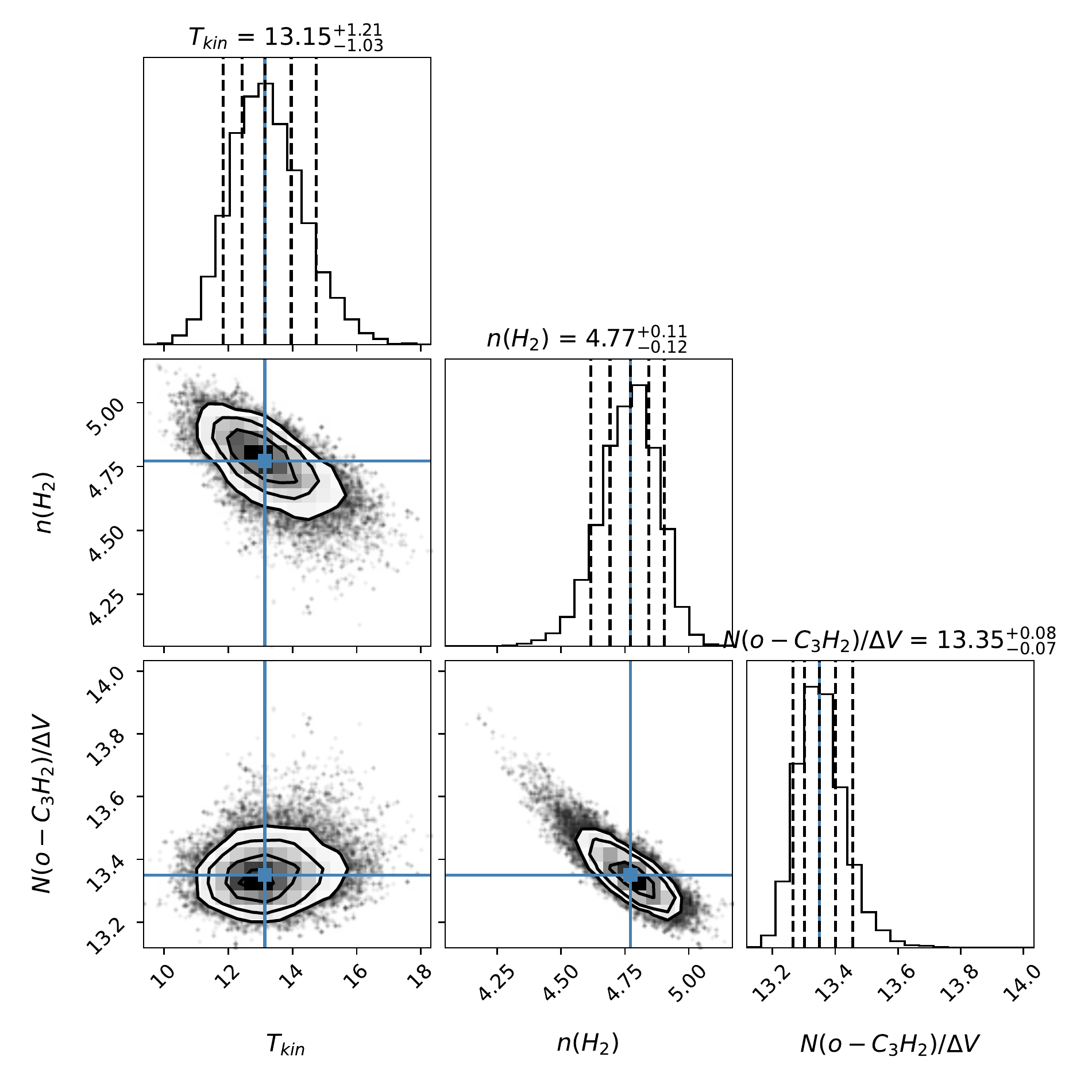}&\\
    \end{tabular}
    \caption{Posterior distribution of $T_{\mathrm{kin}}$, log$_{10}$($n(\mathrm{H_{2}})$), log$_{10}$($N_{\mathrm{mol}}$/$\Delta V$) based on modeling c-C$_{3}$H$_{2}$ lines at the dust peak. All parameters are kept free.}
    \label{fig:corners1}
\end{figure*}  
\begin{figure*}
\begin{tabular}{p{0.45\linewidth}p{0.45\linewidth}}
    \includegraphics[scale=0.4]{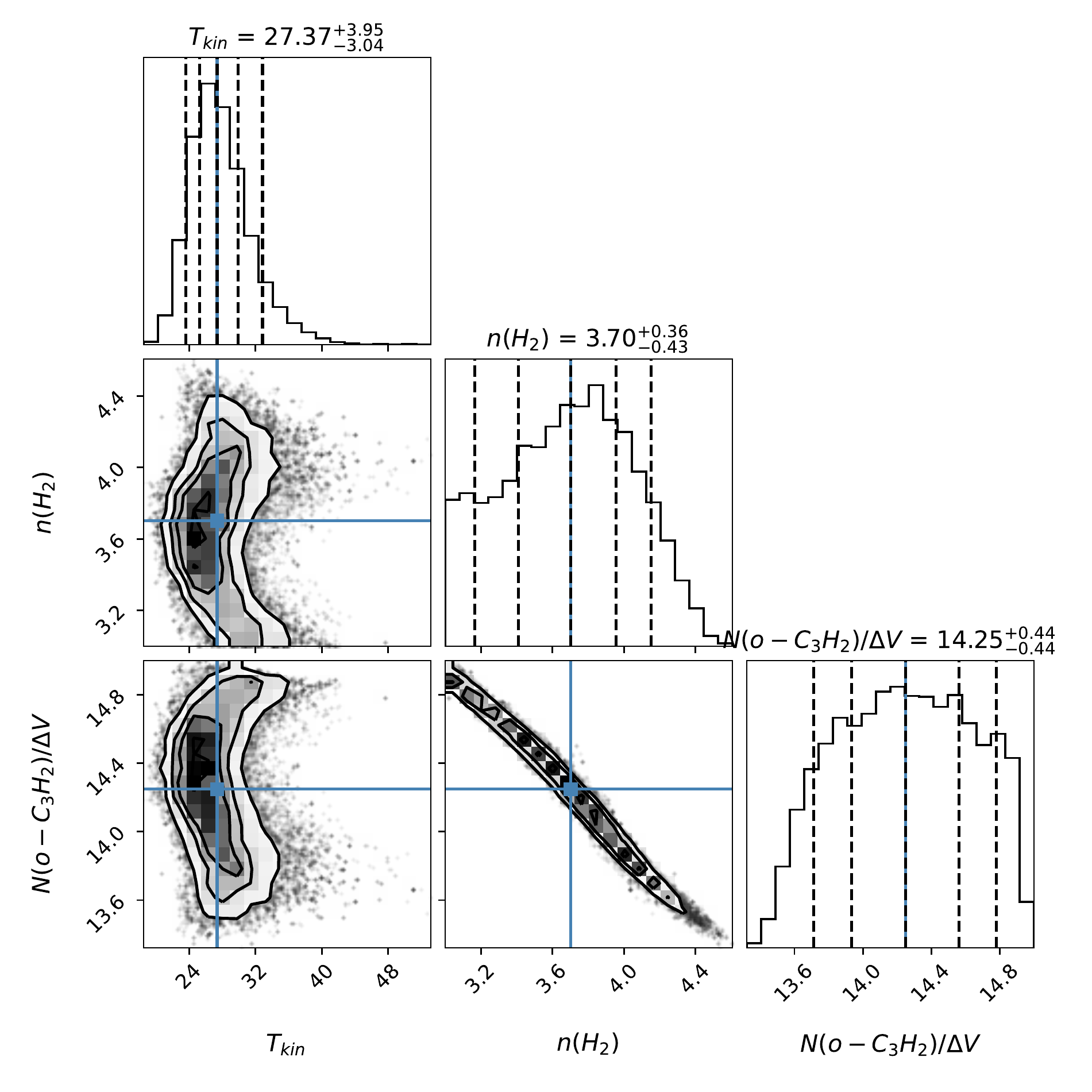}&\includegraphics[scale=0.4]{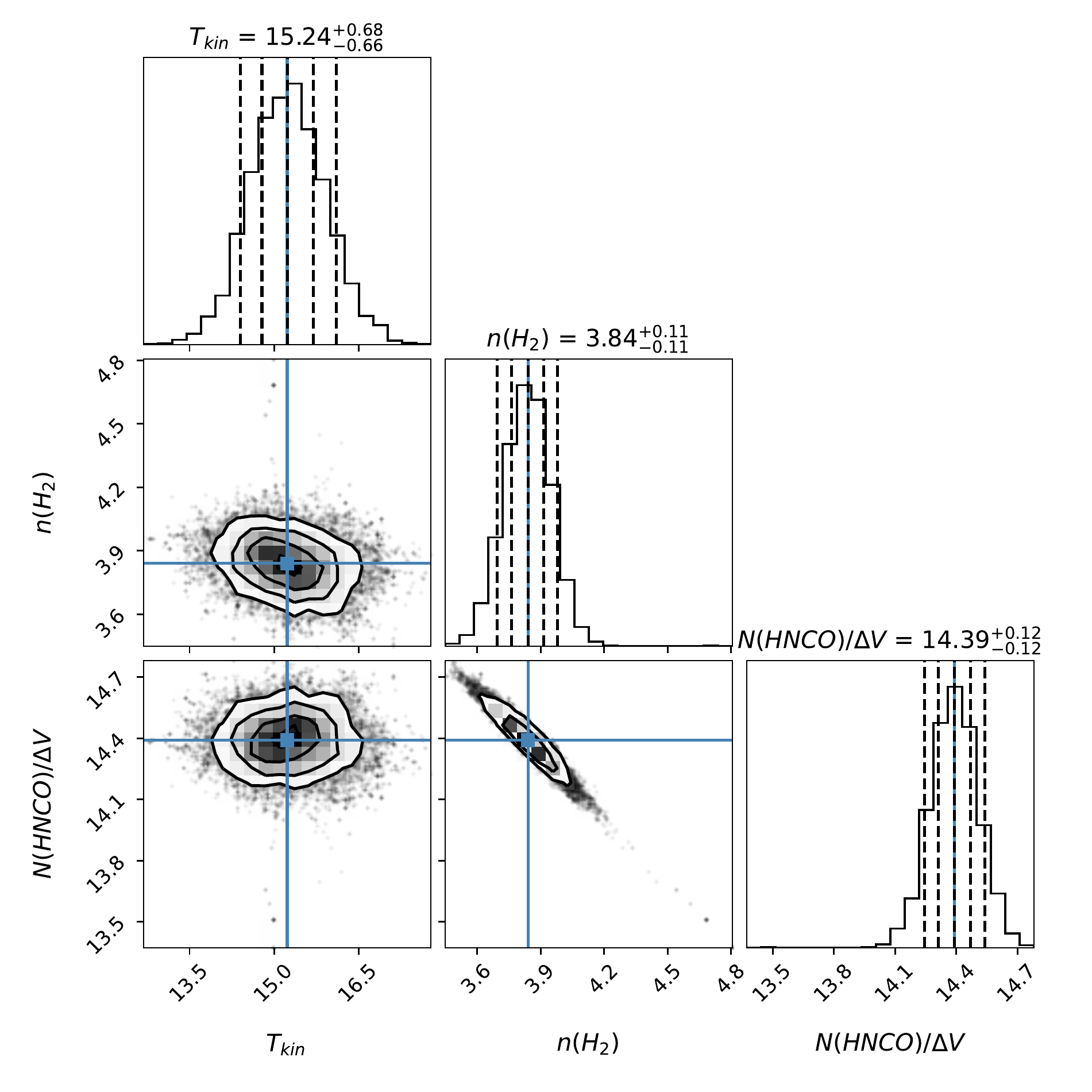}\\
    \end{tabular}
    \caption{Posterior distribution of $T_{\mathrm{kin}}$, log$_{10}$($n(\mathrm{H_{2}})$), log$_{10}$($N_{\mathrm{mol}}$/$\Delta V$) based on modeling c-C$_{3}$H$_{2}$ lines at the o-C$_{3}$H$_{2}$ peak. {\emph{Left:}} All parameters are kept free. {\emph{Right:}} $T_{\mathrm{kin}}$ follows N($\mu$=10 K, $\sigma$=5 K), and the other two parameters are kept free.}
    \label{fig:corners2}
\end{figure*}

\begin{figure*}
\begin{tabular}{p{0.45\linewidth}p{0.45\linewidth}}
    \includegraphics[scale=0.4]{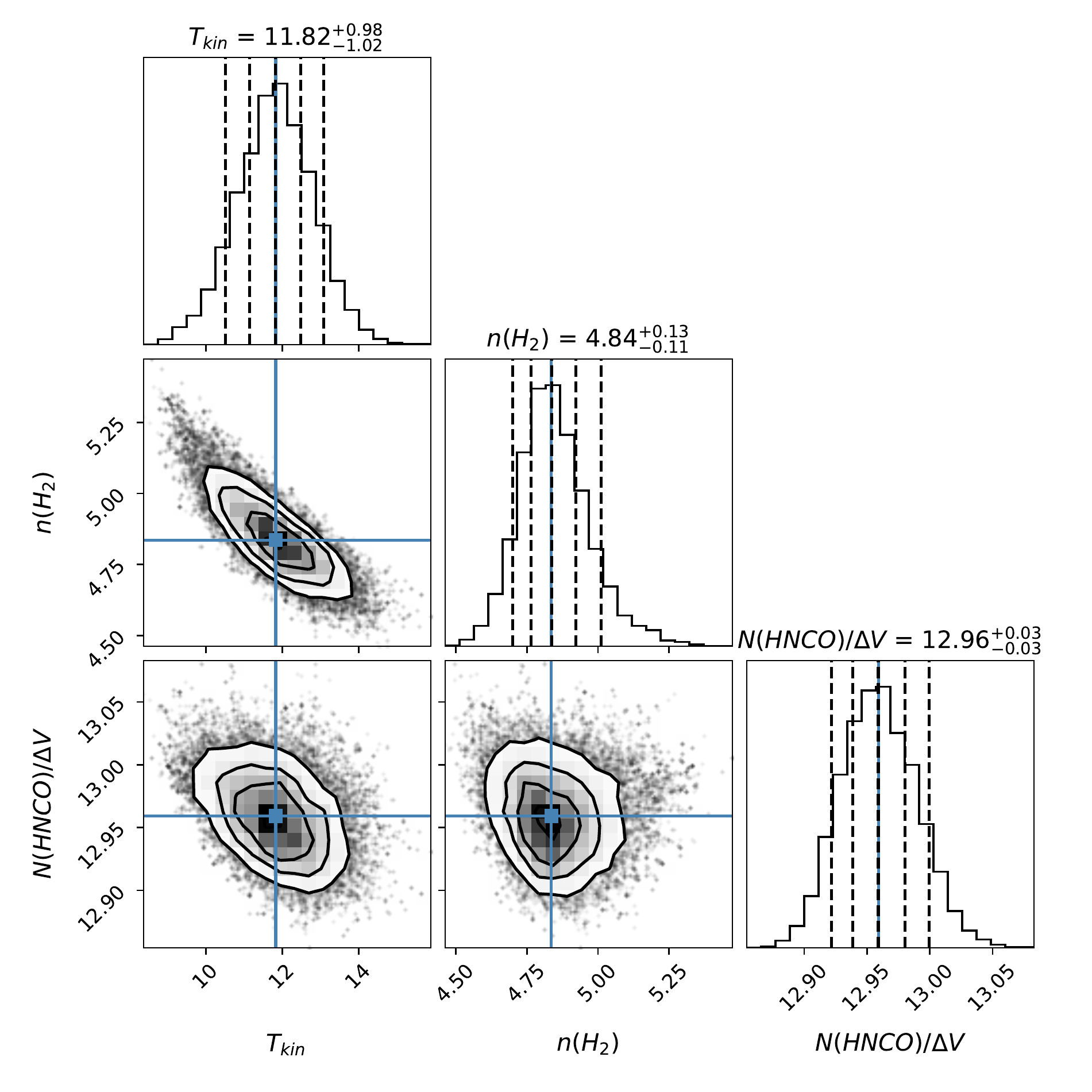}&\includegraphics[scale=0.4]{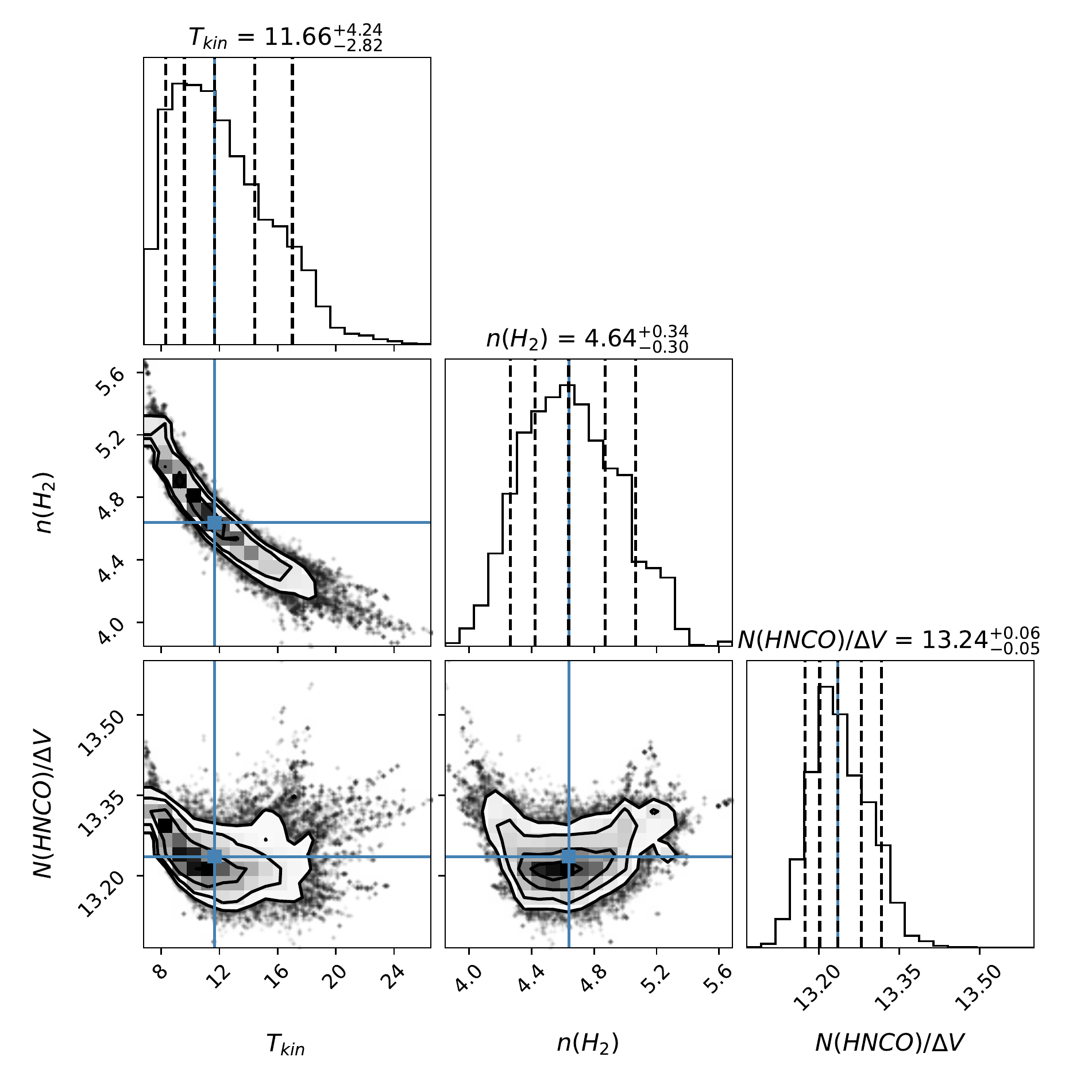}\\
    \end{tabular}
    \caption{Posterior distribution of $T_{\mathrm{kin}}$, log$_{10}$($n(\mathrm{H_{2}})$), log$_{10}$($N_{\mathrm{mol}}$/$\Delta V$) based on modeling HNCO lines at the dust peak and the HNCO peak. $T_{\mathrm{kin}}$ follows N($\mu$=10 K, $\sigma$=5 K) at the dust peak and at the HNCO peak, and the other two parameters are kept free.}
    \label{fig:corners3}
\end{figure*}

\section{Comparison between modelled spectra from LOC and the observational data}
The modelled spectra with LOC for abundance profiles from different epochs (Sec. \ref{sec:loc}) are shown here.  The model spectral cubes were convolved with the respective beam at each frequency, and the spectra were extracted at the position of either dust peak or molecular peaks for comparison with observed lines.  
\begin{figure*}
    \centering
    \hspace{-.9cm}\includegraphics[scale=0.305]{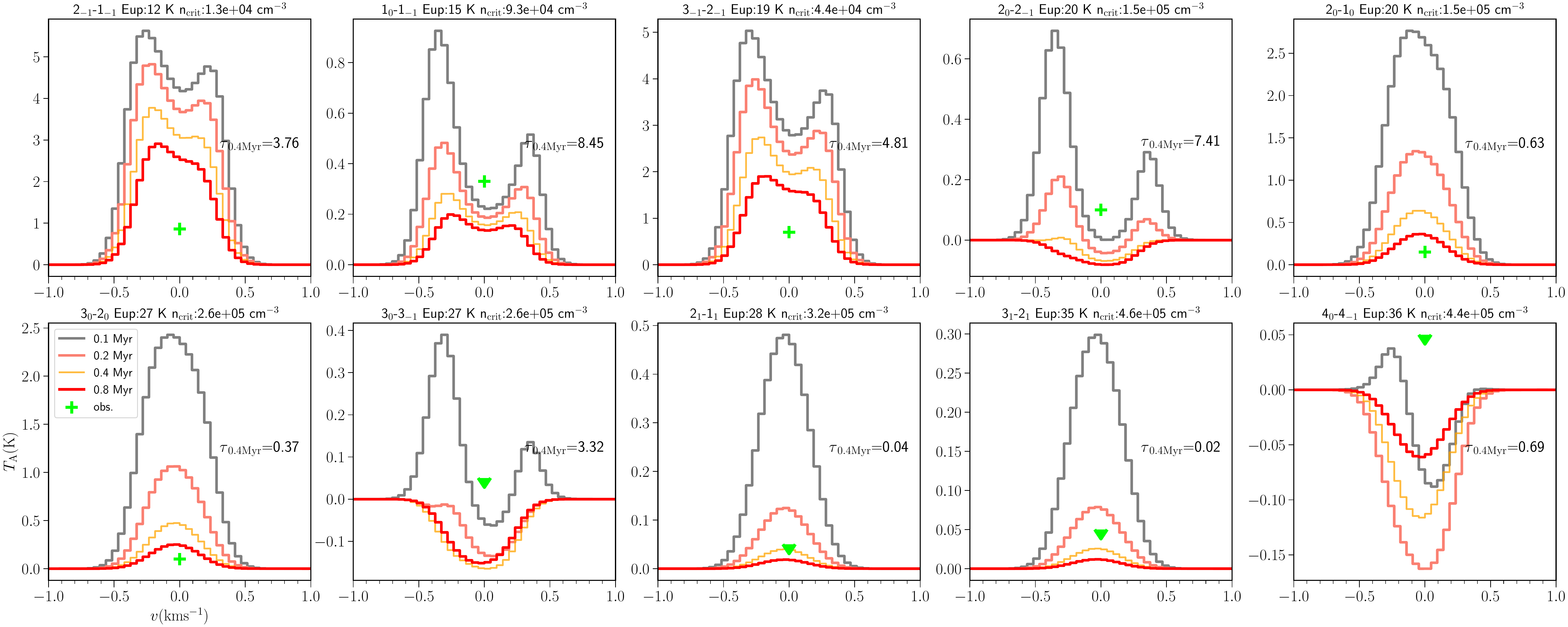}\\
    \hspace{-.9cm}\includegraphics[scale=0.305]{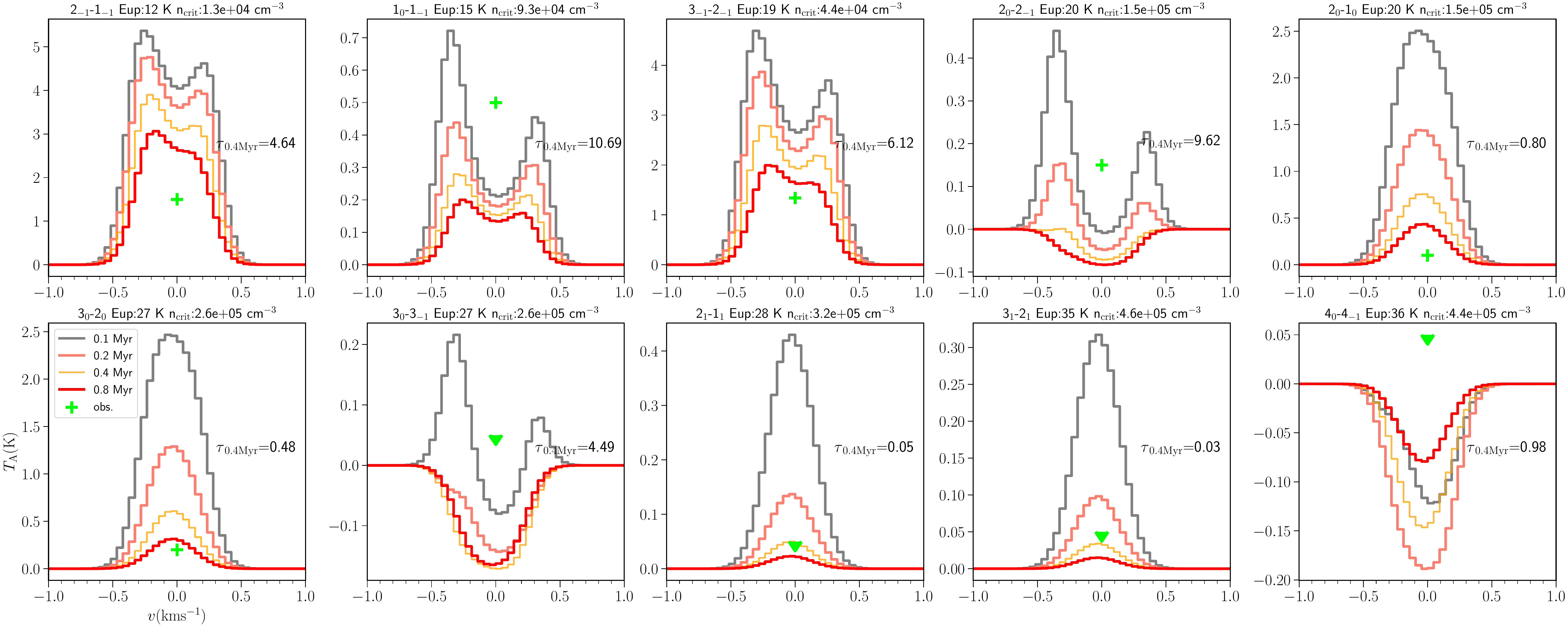}
    \caption{Comparison between modelled spectra and observed CH$_{3}$OH lines, with radial abundance profiles following \citet{vasyunin17} for different epochs in between 0.1-1 Myr. The spectra at the dust peak and at the CH$_{3}$OH peak are shown in upper panel and lower panel, respectively. The optical depth for each modelled line of the 0.4 Myr epoch abundance profile is indicated in each subplot. In the title of each subplot, the quantum numbers, upper energy level and the critical density are shown.}
    \label{fig:vas_ch3oh_model}
\end{figure*}

\begin{figure*}
    \centering
    \hspace{-0.9cm}\includegraphics[scale=0.305]{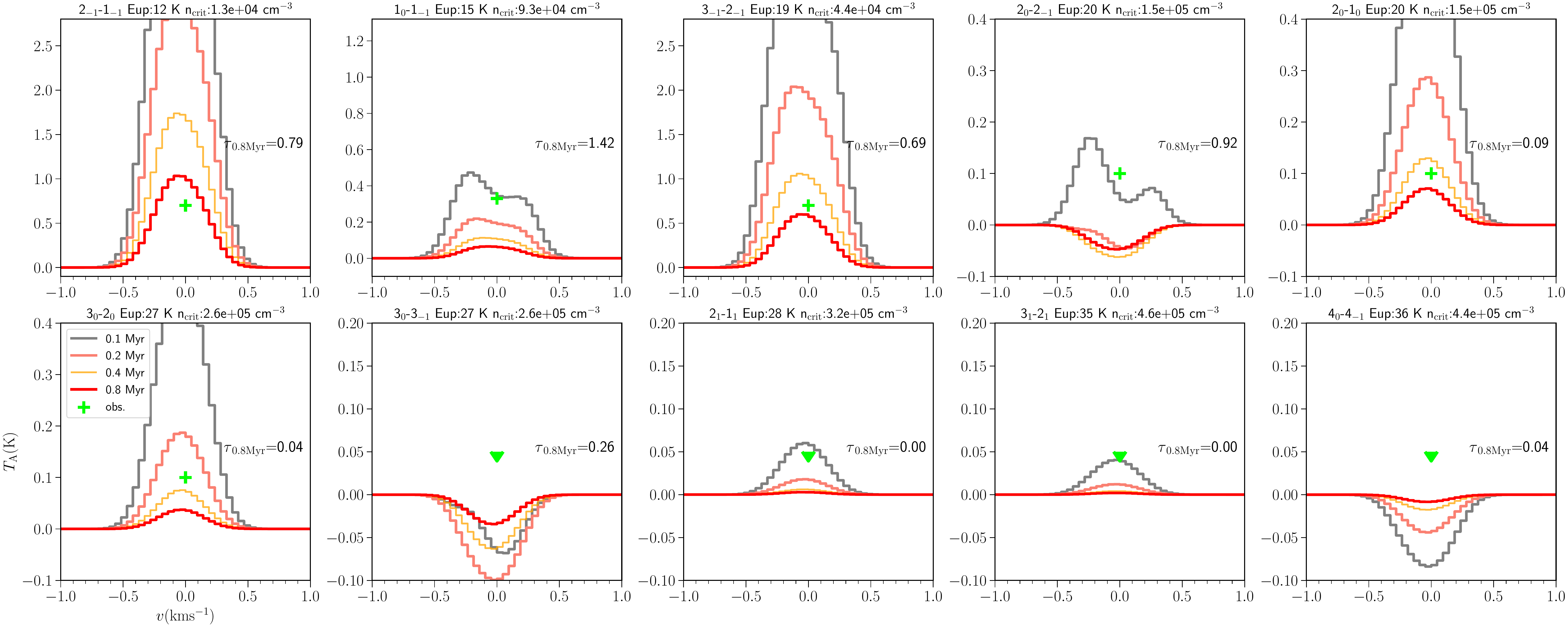}\\
    \hspace{-0.9cm}\includegraphics[scale=0.305]{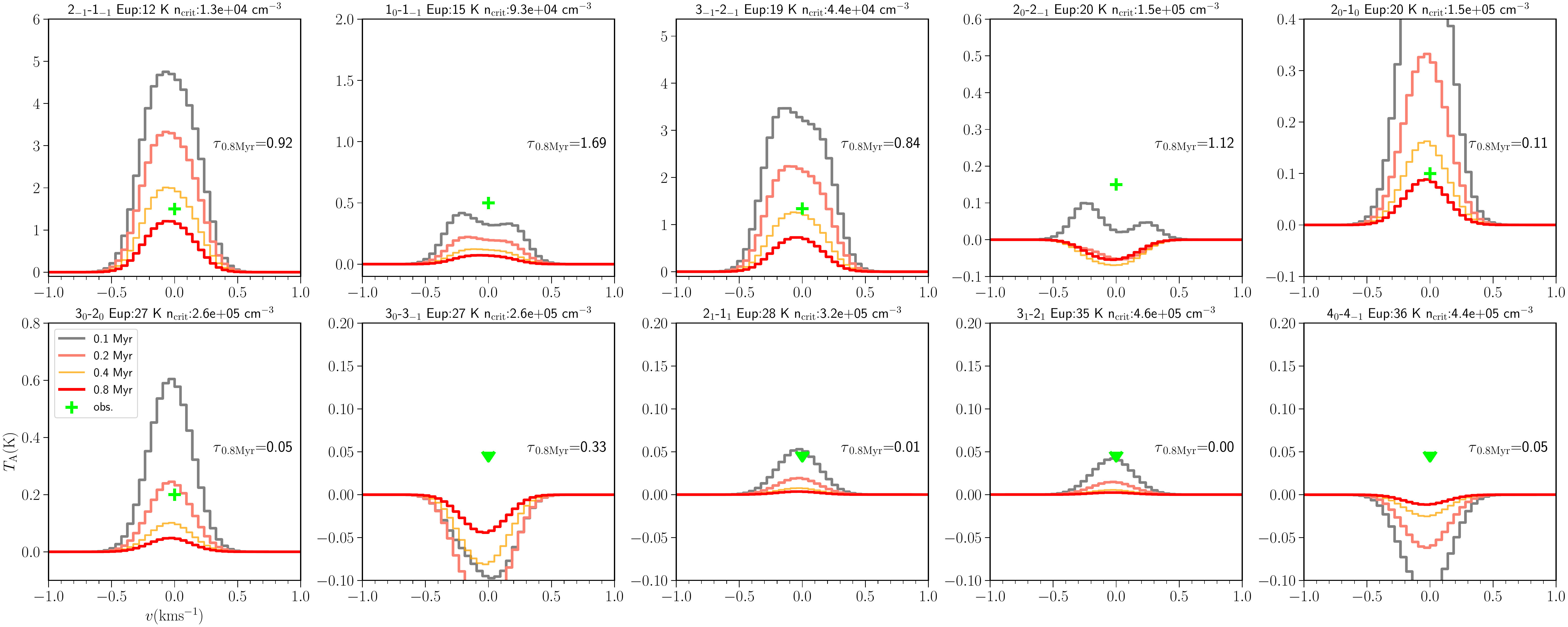}
    \caption{Same as Fig. \ref{fig:vas_ch3oh_model}, but adjusted radial abundance profiles (based on the original abundance profiles in \citet{vasyunin17}. The abundance profile is scaled down by a constant factor of 5.}
    \label{fig:vas_ch3oh_model_adjab}
\end{figure*}

\begin{figure*}
    \centering
     \hspace{-0.9cm}\includegraphics[scale=0.305]{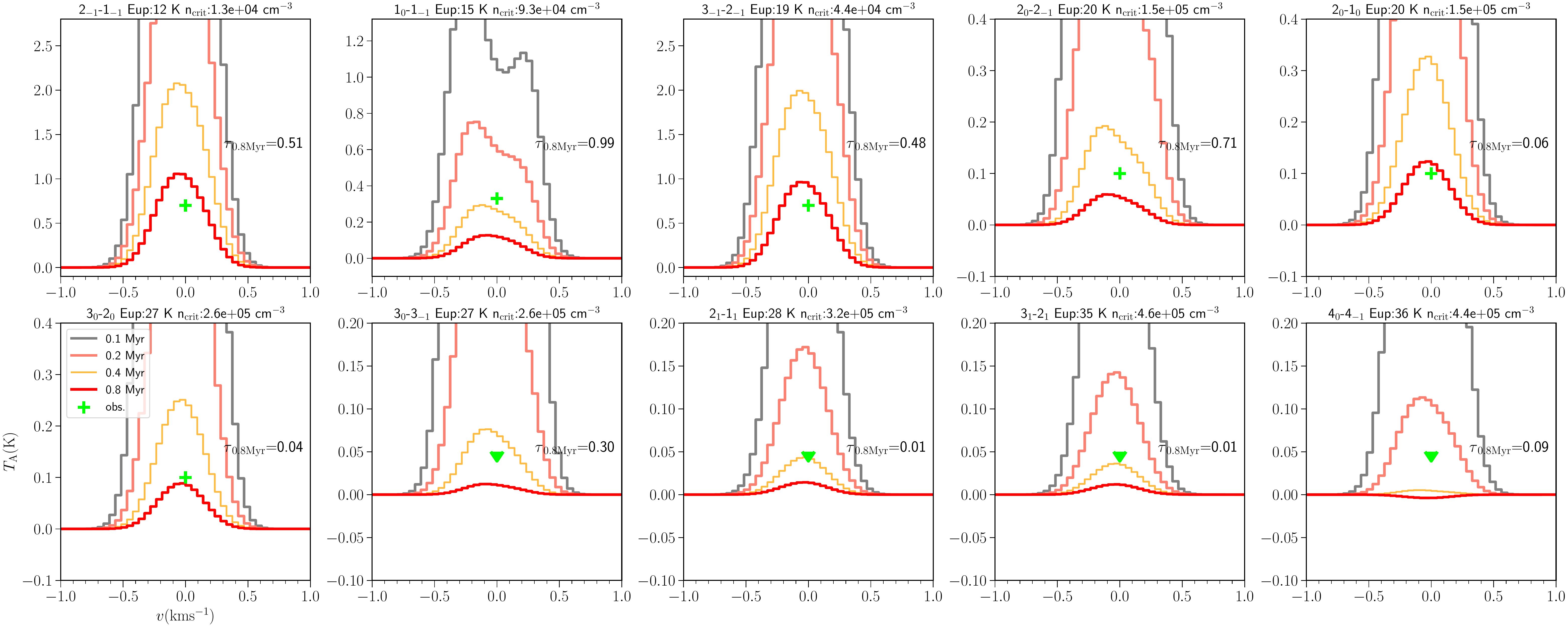}
    \hspace{-0.9cm}\includegraphics[scale=0.305]{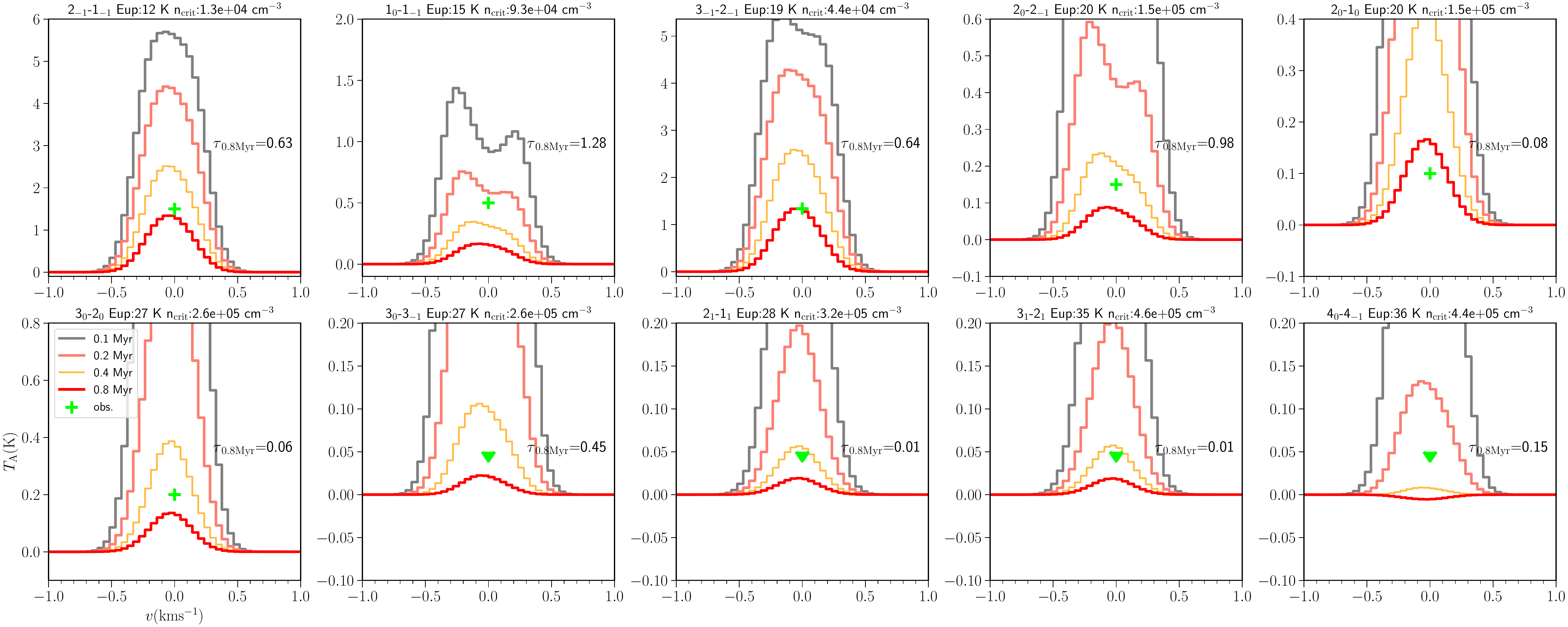}\\
    \caption{Same as Fig. \ref{fig:vas_ch3oh_model}, but adjusted radial abundance profiles (based on the original abundance profiles in \citet{vasyunin17}) and gas density profiles are adopted for the models. The abundance profile is scaled down by a constant factor of 10 and the gas density profile follows the modified radial density profile as in Fig. \ref{fig:dens_bump}.}
    \label{fig:vas_ch3oh_model_adj}
\end{figure*}


\begin{figure*}
    \centering
    \includegraphics[scale=0.35]{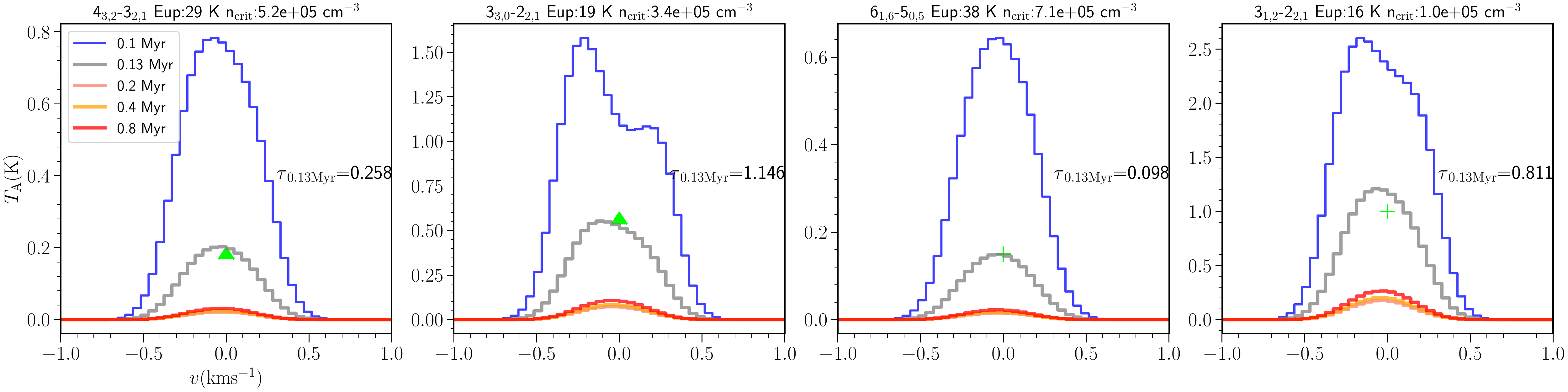}\\
    \includegraphics[scale=0.35]{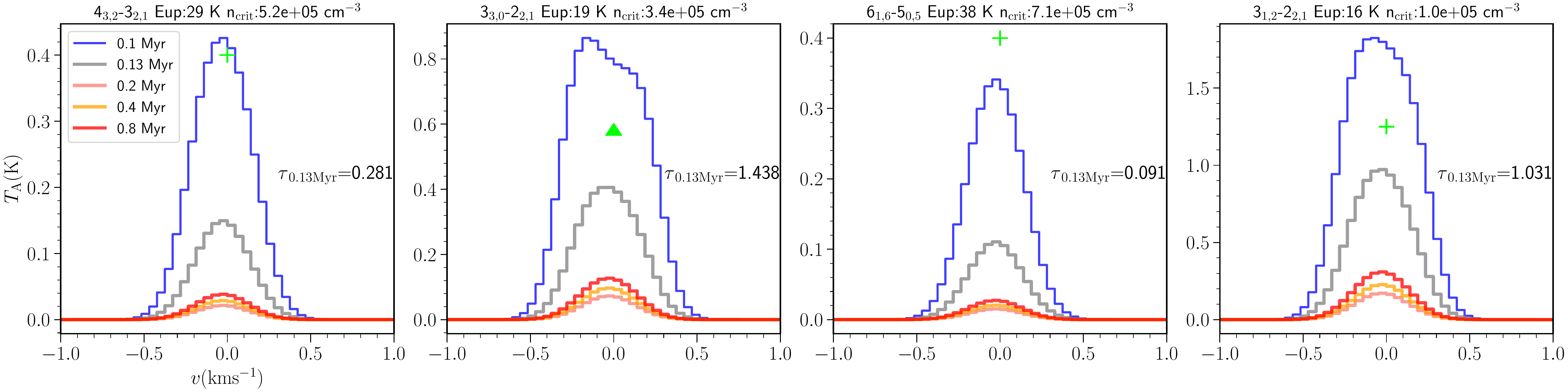}
    \caption{Comparison between modelled spectra and observed c-C$_{3}$H$_{2}$ lines, with radial abundance profiles following \citet{Sipila15} for different epochs in between 0.1-1 Myr. The spectra at the dust peak and at the c-C$_{3}$H$_{2}$ peak are shown in upper panel and lower panel, respectively. The optical depth for each modelled line of the 0.13 Myr epoch abundance profile is indicated in each subplot.}
    \label{fig:olli_c3h2_model}
\end{figure*}

\begin{figure*}
    \centering
    \includegraphics[scale=0.35]{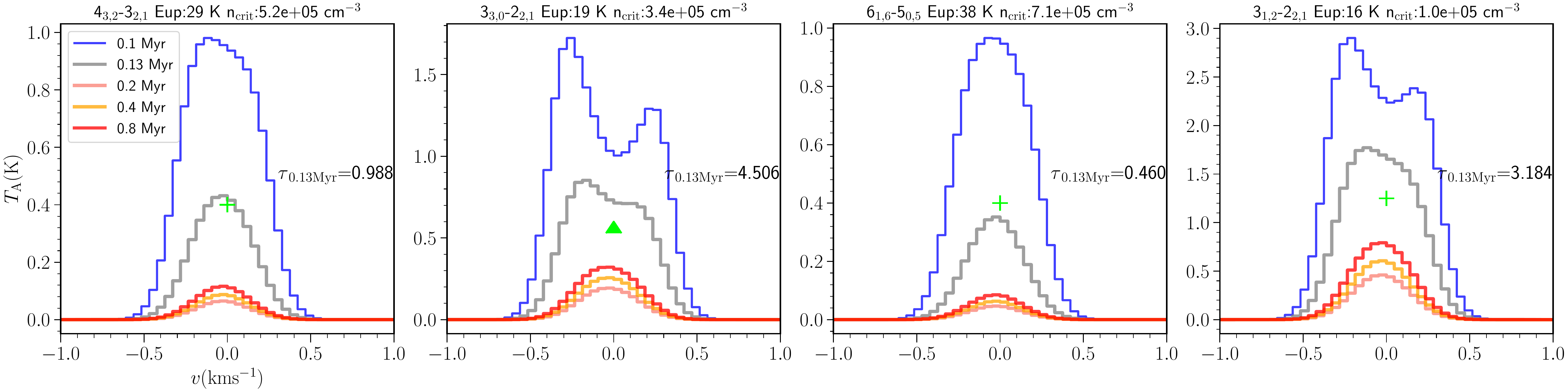}
    \caption{Same as the lower panel of Fig. \ref{fig:olli_c3h2_model}, but with adjusted radial abundance profiles (based on the original abundance profiles in \citet{Sipila15}). The abundance profile is scaled up by a constant factor of 3.}
    \label{fig:olli_c3h2_model_adj}
\end{figure*}

\begin{figure*}
    \centering
    \includegraphics[scale=0.35]{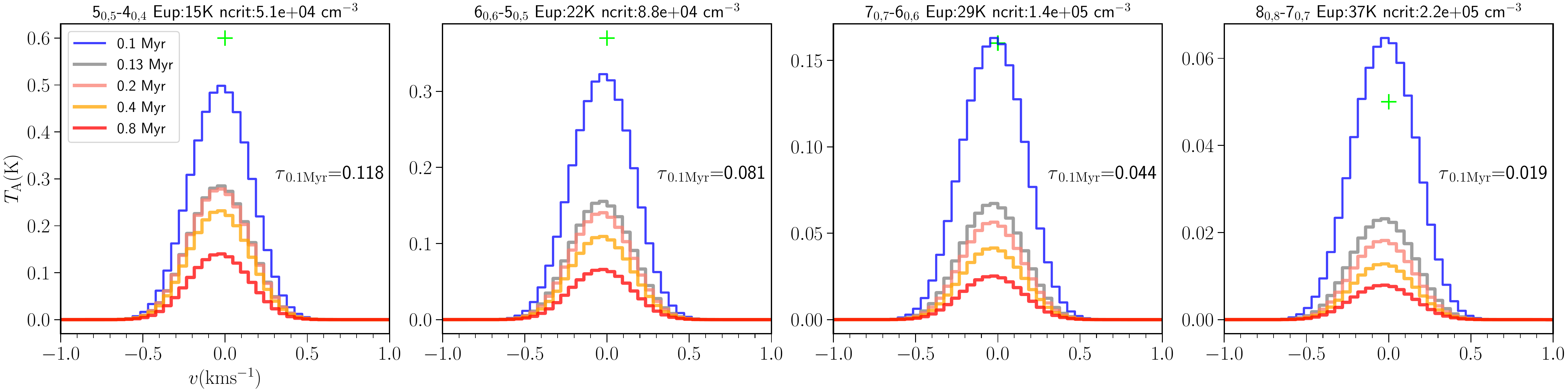}\\
    \includegraphics[scale=0.35]{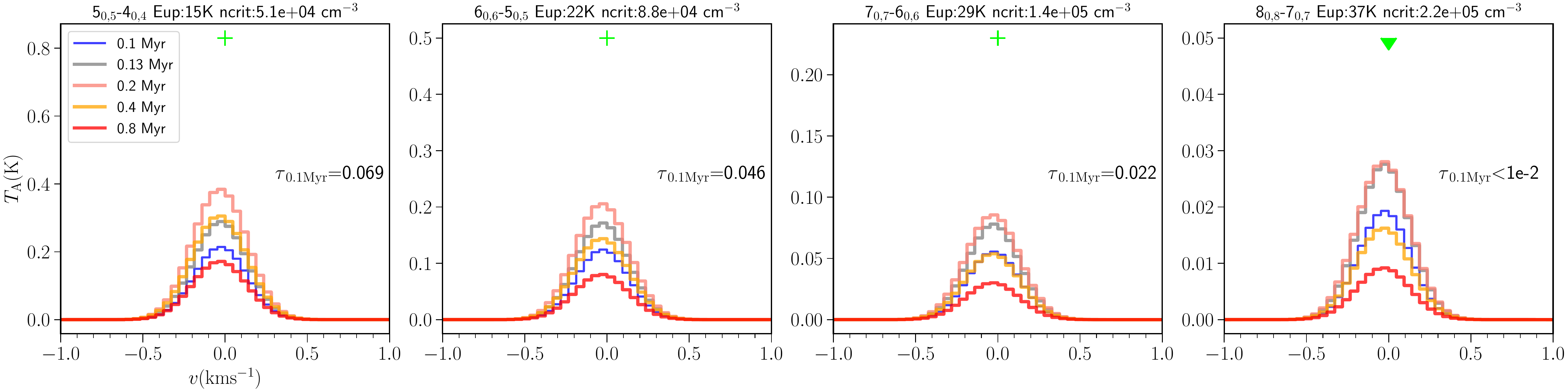}
    \caption{Comparison between modelled spectra and observed HNCO lines, with radial abundance profiles following \citet{Sipila15} for different epochs in between 0.1-1 Myr. The spectra at the dust peak and at the HNCO peak are shown in upper panel and lower panel, respectively. The optical depth for each modelled line of the 0.1 Myr epoch abundance profile is indicated in each subplot.}
    \label{fig:olli_hnco_model}
\end{figure*}

\begin{figure*}
    \centering
   \includegraphics[scale=0.35]{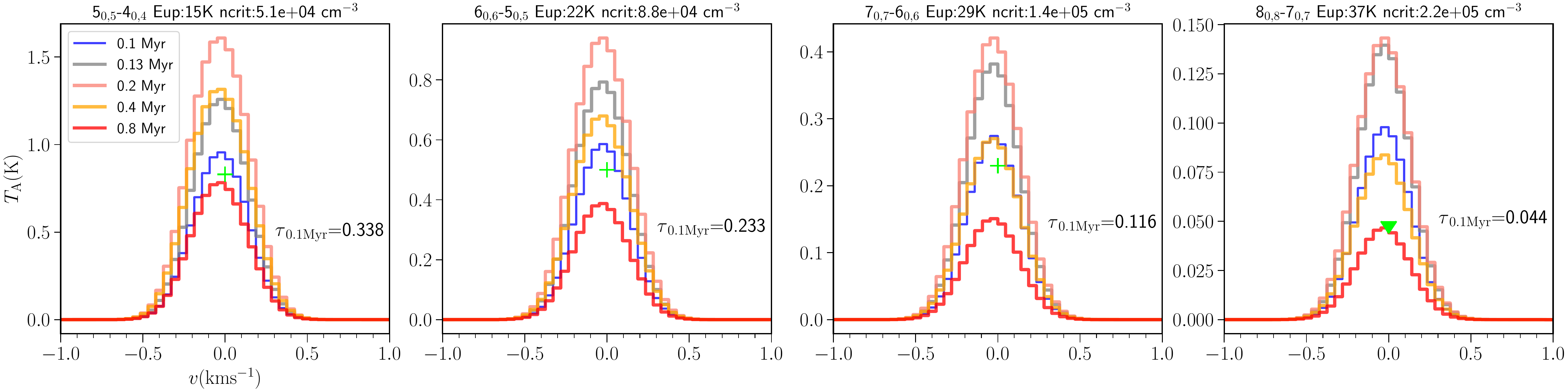}
 \caption{Same as Fig. \ref{fig:olli_hnco_model}, but adjusted radial abundance profiles (based on the original abundance profiles in \citet{Sipila15}). The abundance profile is scaled up by a constant factor of 5.}
  \label{fig:olli_hnco_model_adj}
\end{figure*}

\end{appendix}
\end{document}